\documentclass[useAMS,usenatbib]{mn2e}
\usepackage{float}
\usepackage{amsmath}	% Advanced maths commands
\usepackage{amssymb}	% Extra maths symbols
\usepackage{multicol}   % Multi-column entries in tables
\usepackage{bm}		    % Bold maths symbols, including upright Greek
\usepackage{pdflscape}	% Landscape pages
\usepackage{array}
\usepackage{booktabs}
\usepackage{tabularx}
\usepackage{natbib}
\usepackage{rotating}
\usepackage{subfigure}
\title[Monthly Notices: \LaTeXe\ guide for authors]
  {Photometric redshift estimation of galaxies in the DESI Legacy Imaging Surveys}
\author[C. Li et~al.]
 {Changhua Li$^{1,2,3}$,
 	Yanxia Zhang$^{1,4}$\thanks{Email: zyx@bao.ac.cn},
 	Chenzhou Cui$^{1,3}$\thanks{Email: ccz@bao.ac.cn}, Dongwei Fan$^{1,3}$, \and Yongheng Zhao$^{1}$, Xue-Bing Wu$^{5,6}$, Jing-Yi Zhang$^{1}$, Yihan Tao$^{1,3}$, Jun Han$^{1,3}$,  \and Yunfei Xu$^{1,3}$, Shanshan Li$^{1,2,3}$, Linying Mi$^{1,3}$, Boliang He$^{1,2,3}$, Zihan Kang$^{1,2}$, \and Youfen Wang$^{1,3}$,  Hanxi Yang$^{1,3}$ and Sisi Yang$^{1,3}$\\
 	$^1$ National Astronomical Observatories, Beijing, 100101, China\\
 	$^2$ University of Chinese Academy of Sciences, Beijing 100049, China\\
 	$^3$ National Astronomical Data Center, Beijing 100101, China\\
 	$^4$ CAS Key Laboratory of Optical Astronomy, National Astronomical
 	Observatories, Beijing, 100101, China\\
 	$^5$ Department of Astronomy, School of Physics, Peking University, Beijing 100871, China\\
 	$^6$ Kavli Institute for Astronomy and Astrophysics, Peking University, Beijing 100871, China}
\date{Released 2002 Xxxxx XX}

\begin{document}

\label{firstpage}

\maketitle

\begin{abstract}
The accurate estimation of photometric redshifts plays a crucial role in accomplishing science objectives of the large survey projects. The template-fitting and machine learning are the two main types of methods applied currently. Based on the training set obtained by cross-correlating the DESI Legacy Imaging Surveys DR9 galaxy catalogue and SDSS DR16 galaxy catalogue, the two kinds of methods are used and optimized, such as {\small EAZY} for template-fitting approach and {\small CATBOOST} for machine learning. Then the created models are tested by the cross-matched samples of the DESI Legacy Imaging SurveysDR9 galaxy catalogue with LAMOST DR7, GAMA DR3 and WiggleZ galaxy catalogues. Moreover three machine learning methods ({\small CATBOOST}, Multi-Layer Perceptron and Random Forest) are compared, {\small CATBOOST} shows its superiority for our case. By feature selection and optimization of model parameters, {\small CATBOOST} can obtain higher accuracy with optical and infrared photometric information, the best performance ($\rm MSE=0.0032$, $\sigma_{\rm NMAD}=0.0156$ and $O=0.88$ per~cent) with $g \le 24.0$, $r \le 23.4$ and $z \le 22.5$ is achieved. But {\small EAZY} can provide more accurate photometric redshift estimation for high redshift galaxies, especially beyond the redhisft range of training sample. Finally, we finish the redshift estimation of all DESI DR9 galaxies with {\small CATBOOST} and {\small EAZY}, which will contribute to the further study of galaxies and their properties.
\end{abstract}

\begin{keywords}
Astronomical data bases: catalogues - surveys - Galaxies: general - galaxies: photometry - 
galaxies: distances and redshifts 
\end{keywords}

\section{Introduction} \label{sec:intro}
With the development and use of many large-scale survey telescopes, like the Sloan Digital Sky Survey (SDSS; \citealt{york00}), the Kilo-Degree Survey (KiDS; \citealt{kilo2013}), the Panoramic Survey Telescope \& Rapid Response System (Pan-STARRS; \citealt{panstar2016}), and the Dark Energy Survey (DES; \citealt{DES2005}), multi-band photometries of a large number of galaxies have been obtained. Estimating the photometric redshifts of galaxies is one of the most important steps for probing cosmology and will greatly improve the scientific output of survey telescopes. Since \citet{Puschell1982} first proposed photometric redshift, photometric redshift estimation technologies have been developed rapidly and have become an important issue of astronomical research.

At present, there are many common algorithms for photometric redshift estimation, they can be generally divided into two categories: template-fitting and machine learning methods. The template-fitting methods establish the relationship between synthetic magnitudes (flux) and redshifts by a series of spectral energy distribution templates. The earliest published photometric redshift codes include {\small LePhare} (\citealt{arnouts1999}), {\small BPZ} (Bayesian photometric Redshifts; \citealt{benitez2000}) and {\small HYPERZ} (\citealt{bolzonella2000}). Since then more photometric redshift codes have been devised and published, such as {\small PhotoZ Bayesian} (\citealt{bender2001}), {\small Z-PEG} (\citealt{le2002}), {\small IMPZ} (\citealt{babbedge2004}), {\small ZEBRA} (\citealt{feldmann2006}) and {\small EAZY} (\citealt{eazy2008}). Generally speaking, machine learning algorithms are to find the relation between photometric information and redshifts based on known samples, which is called as training. With this relation, the model is established, it can be used to map the same observable data. Early machine learning approaches include polynomial fitting (\citealt{connolly1995}), and $k$ nearest neighbours (kNN; \citealt{csabai2003}; \citealt*{zhang13}; \citealt{Curran2021}; \citealt{nishizawa2020};). Afterwards, with increasingly powerful computing and storage capabilities, many powerful machine learning algorithms have emerged and were widely applied to photometric redshift estimation, for example, Gaussian process regression (\citealt{way06}; \citealt{way09}; \citealt{bon10}; \citealt{almosallam2016}), kernel regression (\citealt{wang07}), Support Vector Machine (\citealt{jon17}; \citealt{sch17}; \citealt{Jinx2019}), Random Forest (RF; \citealt{car10}; \citealt{sch17}; \citealt{zhang2019}; \citealt{zhou2021}), artificial neural networks (\citealt{collister2007}; \citealt{reis2012}; \citealt{bre13}; \citealt{amaro2019}; \citealt{shuntov2020}) and boosted decision trees (\citealt{Gerdes2010}), {\small XGBOOST} (\citealt{Jinx2019}; \citealt{Li2022}), {\small CATBOOST} (\citealt{Li2022}), etc.

Moreover many works focused on the detailed comparison and application of different codes/algorithms. For instance, \citet{sanchez14} applied most of the existing photo-$z$ codes including some machine learning algorithms and template-fitting methods, to provide photometric redshifts of galaxies from the early data of the Dark Energy Survey. \cite{beck17} explored the performance of machine learning and template-fitting methods in the respect of redshift coverage, extrapolation and photometric errors, and provided the application reference on the choice of machine learning and template-fitting methods. \citet{desprez20} assessed the strengths and weaknesses of machine learning and template-fitting approaches, and found that a combination of both approaches with rejection criteria can outperform any individual method. \citet{schmidt20} compared 12 photo-$z$ algorithms, and indicated the influence of the assumptions underlying each technique on the performance of photometric redshift estimation.

Compared with machine learning methods, the template-fitting approach has three advantages: the first is that it does not require a sample set with known redshifts, the second is that it is not limited by the known sample redshift coverage and can be applied to larger redshifts, and the third is that it provides additional information to physically characterise the objects. For machine learning methods, better accuracy can be obtained in the redshift range of known samples, but the prediction range is also constrained by the redshifts of known samples. Combination of them is a good choice. In this paper, we will employ two kinds of methods to carry out the redshift estimation of the DESI survey DR9 galaxies.

%Although the photometric redshift catalogue of DESI survey galaxies have been released

 %In practice, Template-fitting methods are very simple to implement

\section{Data} \label{sec:data}
The Dark Energy Spectroscopic Instrument (DESI; \citealt{DESI2016a}; \citealt{DESI2016b}) Legacy Imaging Surveys (\citealt{Dey2019}) contain three optical telescopes to provide galaxy and quasar targets for follow-up observation by DESI. They survey the northern Galactic cap at ${\delta} >$ 32$^{\circ}$ and the South Galactic Cap region at ${\delta} \le $34$^{\circ}$ in $g$, $r$, $z$ bands, covering about 14000 deg$^2$, the Legacy Surveys deliver 5$\sigma$ detections of a ``fiducial" $g=24.0$, $r=23.4$ and $z=22.5$ AB mag galaxy with an exponential light profile of half-light radius $r_{\rm half}=0.45$ arcsec. The Data Release 9 (DR9) was released in 2021, which included images and catalogues. In this data release version, using the software package ``TRACTOR" (Lang, Hogg \& Mykytyn 2016), the forced photometries on $W1$ and $W2$ bands of the unWISE coadded images were obtained. 

The spectroscopic data with redshifts are mainly from the Data Release 16 of the Sloan Digital Sky Survey (SDSS; \citealt{york00}), the Data Release 7 of the Large Sky Area Multi-object Fiber Spectroscopic Telescope (LAMOST; \citealt{Cui12,Luo15}), the Data Release 3 of the Galaxy And Mass Assembly (GAMA; \citealt{gama2018}) and the WiggleZ Dark Energy Survey (WiggleZ; \citealt{Parkinson2012}). The cross-matched catalogue of SDSS DR16 (\citealt{lyke20}) and DESI DR9 has been released online~\footnote{https://portal.nersc.gov/cfs/cosmo/data/legacysurvey/dr9/\\north(south)/external/}, which provides SDSS DR16 spectral information (MJD, PLATE, FIBERID). Then we cross match DESI DR9 with the SDSS DR16 galaxy catalogue with $z\_warning=0$ to get spectral redshifts by the same MJD, PLATE and FIBERID. The cross-matched catalogue is called as the DSW sample. Then, we select galaxies with spectral redshifts $z>0$ and $z\_err \le 0.01$ from the LAMOST DR7 galaxy catalogue, those with the redshift quality flag $nQ =4$ and $z \textgreater 0$ from the GAMA DR3 catalogue, and those with the redshift quality flag $Q \ge 4$ and $z \textgreater 0$ from the WiggleZ catalogue, and then all these selected galaxies are respectively cross-matched to the DESI DR9 catalogue by positional cross-match with 1.5 arcsec radius. The nearest sources of all targets are kept, the obtained samples are defined as S\_LAMOST, S\_GAMA, S\_WiggleZ, respectively. After that, we compute the magnitudes corresponding to the model and aperture fluxes in $g, r, z, W1, W2$ bands respectively, and remove the out-of-range sources with $\rm g>24.0$, $\rm r>23.4$, $\rm z>22.5$. We also delete all sources with $maskbits!=0$, because these sources are either in corrupted imaging pixels, or pixels that in the vicinity of bright stars, globular clusters, or nearby galaxies. In this work, the DSW sample is used for training, validation and test, the other samples are applied for external test. The number of each sample is shown in Table~1.

%Finally, the number of the DSW sample is 2 609 100 for training and validation by 5-fold cross validation, and that of the LAMOST, GAMA, WiggleZ sample are 107 178, 126 254, 129 271 for external test, respectively.
\begin{table*}
	\begin{center}
		\caption{Number of objects from each spectroscopic survey that are cross-matched to DESI DR9.}
		\begin{tabular}{rlll}
				\hline
			Survey    & Sample   &No. of objects  & Training, validation, test\\
				\hline
				SDSS & DSW      &2 609 100        & Training, validation, test \\
				LAMOST& S\_LAMOST&107 178         & Test       \\
				GAMA  & S\_GAMA  &126 254         & Test        \\
				WiggleZ   &S\_WiggleZ&129 271     & Test         \\
				\hline
		\end{tabular}		
	\end{center}
\end{table*}

All parameter information about the known samples is indicated in Table~2. The $r$ magnitude distribution and the spectroscopic redshift distribution of the training and test samples are shown in Figure~1. As described in Figure~1, the distribution range of $r$ magnitude for the external test samples (S\_LAMOST, S\_GAMA, S\_WiggleZ) is similar to that of the training sample DSW, and most sources from S\_LAMOST and S\_GAMA are bright while most from S\_WiggleZ are faint. The spectroscopic redshift range for the DSW sample and S\_WiggleZ is from 0 to 2, while that of S\_LAMOST and S\_GAMA is from 0 to 1.

\begin{table*}
	\begin{center}
    		\caption{Parameters in the known samples.}
    		\begin{tabular}{rlll}
    			\hline
    			Parameter&Definition  &Catalogue& Waveband\\
    			\hline
    			release & corresponds to the camera & DESI &\\
    			brickid &Brick ID      &DESI   &\\
    			objid & Source ID & DESI & \\
    			ra &Right ascension in decimal degrees &DESI &\\
    			dec &Declination in decimal degrees    &DESI &\\
    			$\rm flux\_g$ &Model flux in $g$ band  &DESI &Optical band\\
    			$\rm flux\_r$ &Model flux in $r$ band  &DESI &Optical band\\
    			$\rm flux\_z$ &Model flux in $z$ band  &DESI &Optical band\\
    			$\rm flux\_w1$   &Model flux in $W1$ band &DESI &Infrared band\\
    			$\rm flux\_w2$   &Model flux in $W2$ band &DESI &Infrared band\\
    			$g$ &Model magnitude in $g$ band  &DESI &Optical band\\
    			$r$ &Model magnitude in $r$ band  &DESI &Optical band\\
    			$z$ &Model magnitude in $z$ band  &DESI &Optical band\\
    			$W1$   &Model magnitude in $W1$ band  &DESI &Infrared band\\
    			$W2$   &Model magnitude in $W2$ band  &DESI &Infrared band\\
    			$\rm flux\_ivar\_g$ &Inverse variance of flux\_g  &DESI &Optical band\\
    			$\rm flux\_ivar\_r$ &Inverse variance of flux\_r   &DESI &Optical band\\
    			$\rm flux\_ivar\_z$ &Inverse variance of flux\_z  &DESI &Optical band\\
    			$\rm flux\_ivar\_w1$ &Inverse variance of flux\_w1  &DESI &Optical band\\
    			$\rm flux\_ivar\_w2$ &Inverse variance of flux\_w2  &DESI &Optical band\\
    			$\rm apflux\_g$   &Aperture fluxes in $g$ band  &DESI  &Optical band\\
    			$\rm apflux\_r$   &Aperture fluxes in $r$ band  &DESI  &Optical band\\
    			$\rm apflux\_z$   &Aperture fluxes in $z$ band  &DESI  &Optical band\\
    			$\rm apflux\_w1$   &Aperture fluxes in $w1$ band  &DESI  &Infrared band\\
    			$\rm apflux\_w2$   &Aperture fluxes in $w2$ band  &DESI  &Infrared band\\    	
    			$\rm ap\_g\_(1\sim 8)$   &Aperture magnitude in $g$ band  &DESI  &Optical band\\
    			$\rm ap\_r\_(1\sim 8)$   &Aperture magnitude in $r$ band  &DESI  &Optical band\\
    			$\rm ap\_z\_(1\sim 8)$   &Aperture magnitude in $z$ band  &DESI  &Optical band\\
    			$\rm ap\_W1\_(1\sim 5)$   &Aperture magnitude in $W1$ band  &DESI  &Infrared band\\
    			$\rm ap\_W2\_(1\sim 5)$   &Aperture magnitude in $W2$ band  &DESI  &Infrared band\\    		
    			$\rm z\_spec$ &Spectral redshift  &SDSS, LAMOST&\\
    			 &  & GAMA, WiggleZ&\\
    			\hline
    		\end{tabular}
    		\bigskip
    	\end{center}
    \end{table*}

\begin{figure*}
    		\centering
		\includegraphics[bb=84 239 531 558,height=6cm]{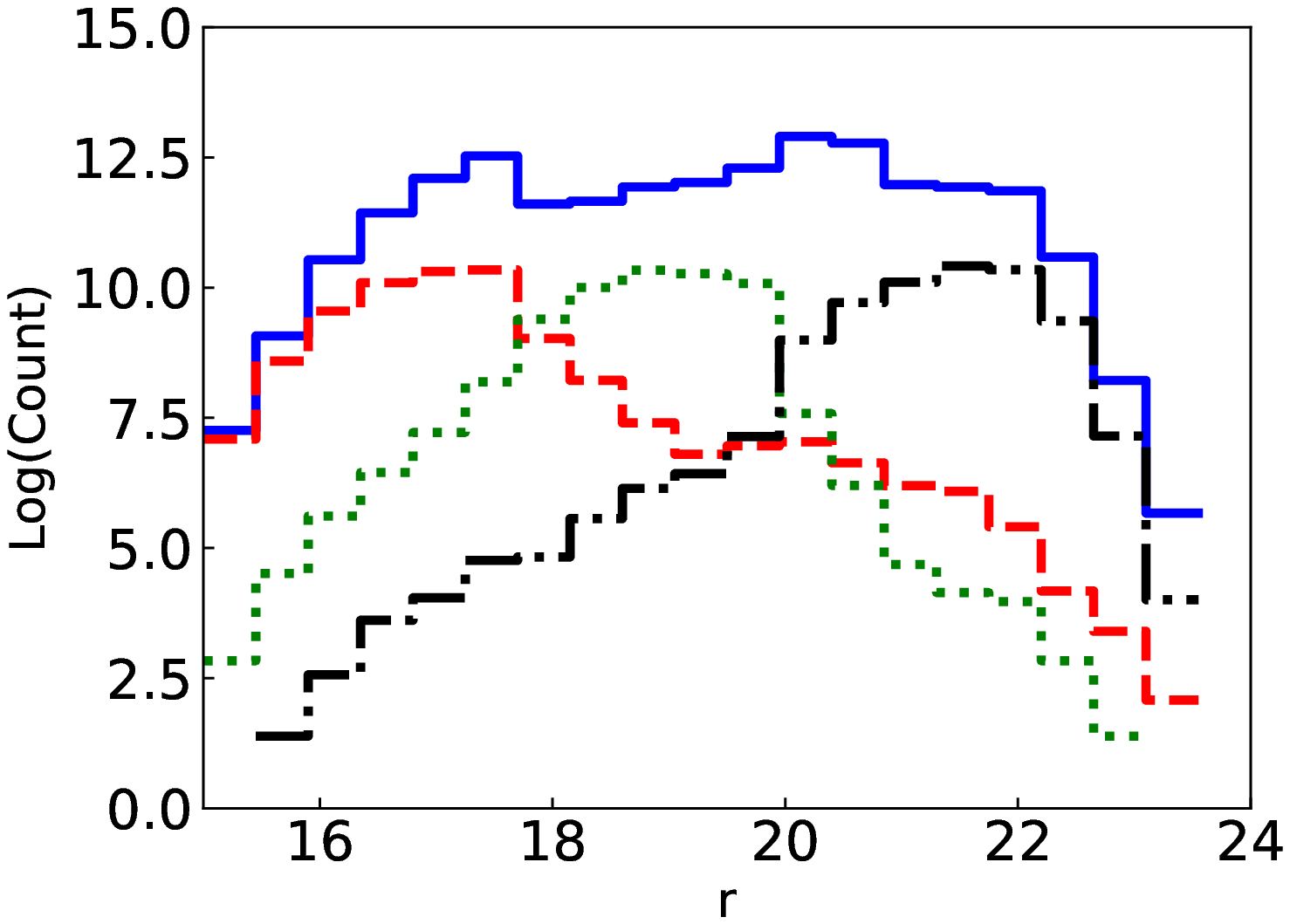}
    		\includegraphics[bb=84 239 531 558,height=6cm]{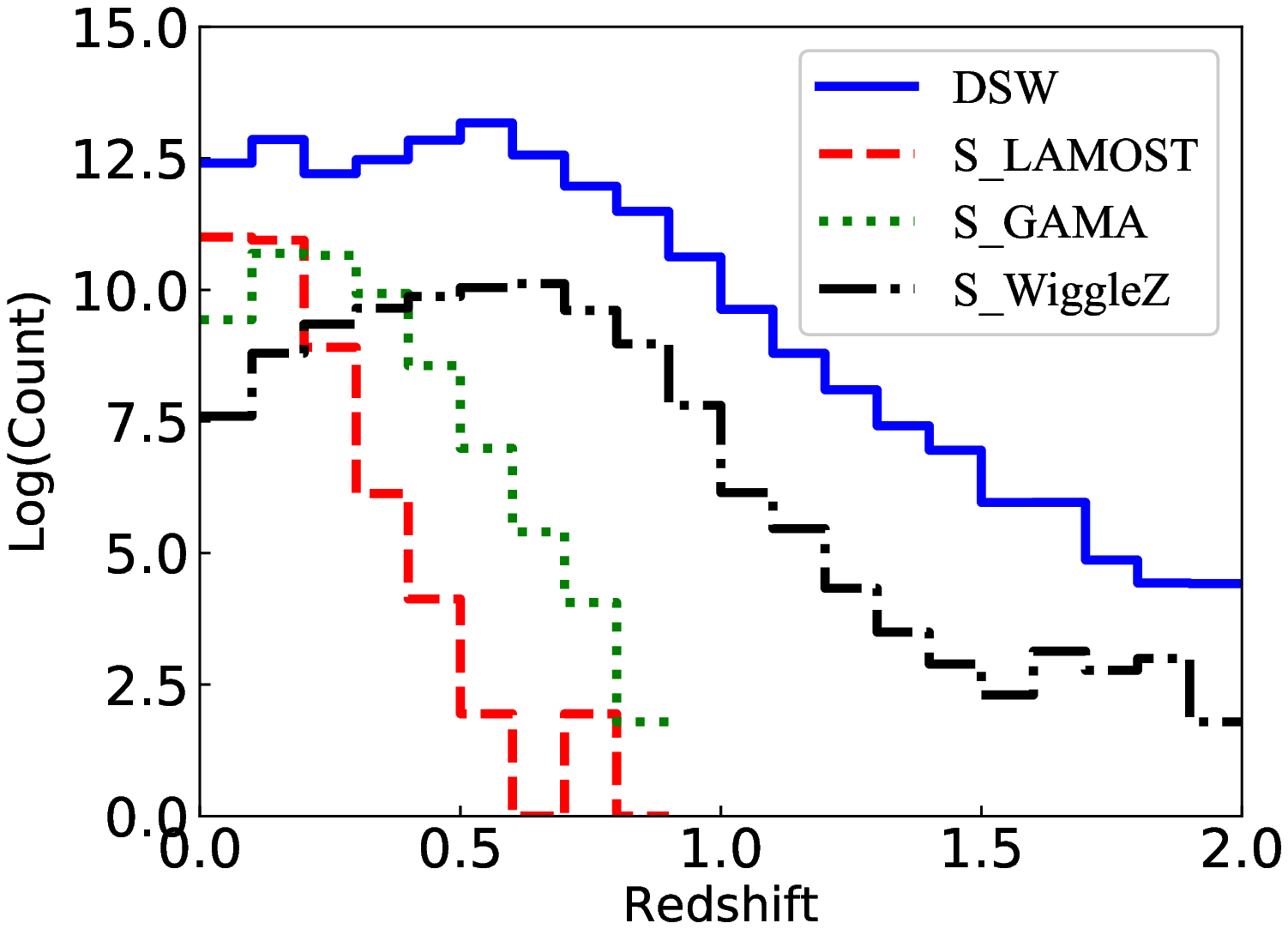}   	
    		\caption{Left panel: the number of galaxies as a function of $r$ magnitude for the training (DSW) and test samples (S\_LAMOST, S\_GAMA, S\_WiggleZ); right panel: the number of galaxies as a function of spectroscopic redshifts for the training (DSW) and test samples (S\_LAMOST, S\_GAMA, S\_WiggleZ).}	
    		\label{fig1}
\end{figure*}

\section{Method} \label{sec:method}
\subsection{{\small EAZY} approach}
{\small EAZY} (\citealt{eazy2008}) is a public photometric redshift code based on template-fitting algorithm, which combines features from various existing programs, such as the possibility of fitting linear combination of templates as done in GREGX (\citealt{ruknick2001}; \citealt{ruknick2003}), and a user-friendly interface based on {\small HyperZ} (\citealt{bolzonella2000}). %The default template set of {\small EAZY} are from semi-analytical models.\\
The basic algorithm of {\small EAZY} is to establish a redshift grid based on the SED templates, find the best-fitting template spectrum in each redshift grid, which minimizes the value of $\chi^2$, and no priors are needed.
\begin{equation}
\chi^2_{z,i} = \sum_{j=1}^{N_{filt}}\frac{(T_{z,i,j} - F_j)^2}{(\delta F_j)^2} \\
\end{equation}
where $N_{filt}$ is the number of filters, $T_{z,i,j} $ is the synthetic flux of template $i$ in filter $j$ for redshift $z$, $F_j$ is the observed flux in filter $j$, and $\delta F_j$ is the uncertainty in $F_j$.
Then, in {\small EAZY}, the linear combination of templates is implemented. In this improved algorithm, it finds the best-fitting coefficients $\alpha$ of all combined templates, so that
\begin{equation}
	T_z = \sum_{i=1}^{N_{temp}} \alpha _i T_{z,i}
\end{equation}
This improvement is significant for accuracy, but more computation time is required.
Meanwhile, for each estimation, {\small EAZY} calculates a parameter $Q_{\rm z}$ (see Equation~3), which demonstrates photometric redshift estimation quality. In general, the estimated redshift with $Q_{\rm z}< 1$ is considered as reliable.
\begin{equation}
Q_{\rm z} = \frac{\chi^2}{N_{\rm filter} - 3 } \frac{z^{\rm 99}_{\rm up} - z^{\rm 99}_{\rm lo}}{p_{\Delta{z}=0.2}}
\end{equation}
In Equation~3, $\chi^2$, $N_{\rm filter}$ is the same as Equation~1, $p_{\Delta{z}}$ represents the fraction of the total integrated probability that lies within $\pm \Delta{z}$ of the estimated redshifts (\citealt{eazy2008}).

\citet{yang2014} estimated photometric redshifts of the Hawaii-Hubble deep field-north(H-HDF-N) survey catalogue by {\small EAZY}. \citet{chen2018} adopted {\small EAZY} to measure the redshifts of the XMM-Newton point source catalogue. The good performance is obtained in these researches. Thus, in this paper, we adopt {\small EAZY} as the representative of template-fitting method to predict the redshifts of DESI DR9 galaxies.

\subsection{{\small CATBOOST}}
Gradient Boosted Decision Trees (GBDT, \citealt{Friedman2001}) are well fit for classification and regression tasks. {\small CATBOOST} \citep{catboost2018}, as a member of the family of GBDT, is a high-performance open source algorithm. {\small CATBOOST} applies boosting method to build strong classifiers or regressors by means of learning multiple weak classifiers or regressors. Comparing to GBDT, {\small CATBOOST} adopts Oblivious Decision Trees (ODT) to build decision trees, which are full binary trees, furthermore, all non-leaf nodes of ODT will have the same splitting criteria. This design is helpful to speed up the score and avoid over fitting. It has good characteristics of supporting categorical features and missing value in training data without additional preprocessing, which improves its usage efficiency. In addition, it achieves easily good performance with default model parameters.

\citet{Li2022} compared the photometric redshift performance of {\small CATBOOST}, {\small XGBOOST} (\citealt{Chen2016}), and Random Forest (RF; \citealt{Bre01}), and {\small CATBOOST} showed the best performance and the fastest training speed. Therefore, we adopt {\small CATBOOST} as the main machine learning method to predict redshifts of DESI DR9 galaxies. We perform all computation of this work provided by National Astronomical Data Centre (NADC) \citep{Li2017}.

\section{Photometric redshift estimation} \label{sec:performance}
\subsection{Metrics}
The performance of different algorithms on photometric redshift estimation may be judged by different metrics, for example, the residual between the spectroscopic and photometric redshifts, $\Delta \mathrm{ z}=\rm z_{\rm spec} - \rm z_{\rm photo}$, the mean absolute error (${\rm MAE}$) and the mean squared error (${\rm MSE}$). They are computed by the following equations:
\begin{equation}
	\mathrm MAE = \frac{1}{n}\sum_{i=0}^{n-1} | \Delta \mathrm{ z} | \\
\end{equation}

\begin{equation}
   \mathrm MSE =\frac{1}{n}\sum_{i=0}^{n-1} ( \Delta \mathrm{ z} )^2 \\
   \end{equation}
%where $z_{i}$ is the true redshift, $\widehat{z_{i}}$ is the predicted redshift value and $n$ is the sample size.

The fraction of test sample that satisfies $\mid \Delta \rm z \mid <e$ is usually used to evaluate the redshift estimation, where $e$ is a given residual threshold (\citealt{sch17} and references therein). In general, the redshift normalized residuals $\rm \Delta{\mathrm{z(norm)}}$ are applied, and here $e=0.3$ is adopted.
\begin{equation}
\rm{{\Delta}z(norm)} = \frac{z_{\rm spec} - z_{\rm phot}}{ 1 + z_{\rm spec}}  \\
\end{equation}

\begin{equation}
\rm{\delta_{0.3}} = \frac{\rm N_{\left|{\Delta}z(norm) \right| < 0.3}}{\rm N_{total}}\\
\end{equation}

Four additional metrics are defined as follows and used for performance evaluation of different machine learning methods: bias (the average separation between prediction and true values), the normalized standard deviation of the photometric redshifts and the spectroscopic redshifts, the normalized median absolute deviation ($\sigma_{\rm NMAD}$) and the outlier fraction ($O$; \citealt{Henghes2021}; \citealt{Curran2021}).

\begin{equation}
\mathrm{Bias}=<z_{\rm spec} - z_{\rm phot}> \\
\end{equation}

\begin{equation}
\sigma_{{\Delta}\mathrm {z(norm)}} =\sqrt{\frac{1}{n}\sum_{i=0}^{n-1} (\rm  {\Delta}z(norm) - \overline{{\Delta}z(norm)} )^2 }\\
\end{equation}

\begin{equation}
\sigma_{\mathrm {NMAD}} = 1.48 \times \rm{median} \left| {\Delta}\mathrm {z(norm)} \right|\\
\end{equation}

%\begin{equation}
%\sigma_{\rm rms} =\sqrt{\frac{1}{n}\sum_{i=0}^{n-1} (\rm  {\Delta}z(norm))^2 }\\
%\end{equation}

\begin{equation}
Outlier\,\, fraction(O) = \frac{\rm N_{\left|{\Delta}z(norm) \right| > 0.15}}{N_{\rm total}}
\end{equation}

\subsection{Photometric redshift estimation by {\small EAZY}}
We estimate photometric redshifts of all galaxies from training set by {\small EAZY}. For {\small EAZY}, the default v1.3 template set is adopted, which includes 9 templates. The parameter $N\_MIN\_COLORS$ is set to 4, which means that the number of filters with values is at least 4, and the other parameters are set as the default. Since each telescope has the respective filter response curves, the data in the south sky or north sky need to be estimated separately. Table~3 shows the performance of template-fitting approach for the south and north subsamples of the DSW sample, and Table~4 indicates the performance of template-fitting approach with $Q_{\rm z}<1$. The performance metrics are greatly improved, comparing the metrics in Table~4 with those in Table~3. If meeting the condition $Q_{\rm z}<1$, the estimated redshifts are more reliable. The scatter figure and $\Delta$z(norm) distribution of estimated photometric redshifts and spectroscopic redshifts are indicated in Figure~2.

 \begin{table*}
 	\begin{center}
 		\caption[]{The performance of photometric redshift estimation with {\small EAZY}.\label{tab:fsel}}
 		\begin{tabular}{rccccccc}
 			\hline
 			Subsample   &     MSE   &MAE   & $Bias$ & $\sigma_{\rm NMAD}$&$\sigma_{\Delta \rm {z(norm)}}$&$\delta_{0.3}$(\%)&$O$(\%)\\
 			\hline
			DSW\_south  &    0.0301 &0.0719&-0.0217 &0.0243              &0.1128                       &98.79             &5.12 \\
 			DSW\_north  &    0.0341 &0.0621&0.0104  &0.0224              &0.1283                       &99.18             &3.43 \\ 			
 			\hline
 		\end{tabular}
 	\end{center}
 \end{table*}

  \begin{table*}
  	\begin{center}
  		\caption[]{The performance of photometric redshift estimation by {\small EAZY} when $Q_{\rm z}<1$.\label{tab:fsel}}
  		\begin{tabular}{rcccccccc}
  			\hline
  			Subsample  &   MSE &MAE   & $Bias$ &$\sigma_{\rm NMAD}$&$\sigma_{\Delta \rm {z(norm)}}$&$\delta_{0.3}$(\%)&$O$(\%)& $Q_{\rm z}>1$(\%)\\
  			\hline
  			DSW\_south  & 0.0088&0.0521&-0.0149 &0.0219             &0.0539                       &99.73             &2.06   & 10.43\\
  			DSW\_north  & 0.0085&0.0464&0.0044  &0.0209             &0.0574                       &99.83             &1.77   & 6.56 \\ 			
  			\hline
  		\end{tabular}
  	\end{center}
  \end{table*}

     \begin{figure*}
     	\centering    	
     		\subfigure[For the subsample DSW\_south]{
     			\includegraphics[height=5cm,width=7cm]{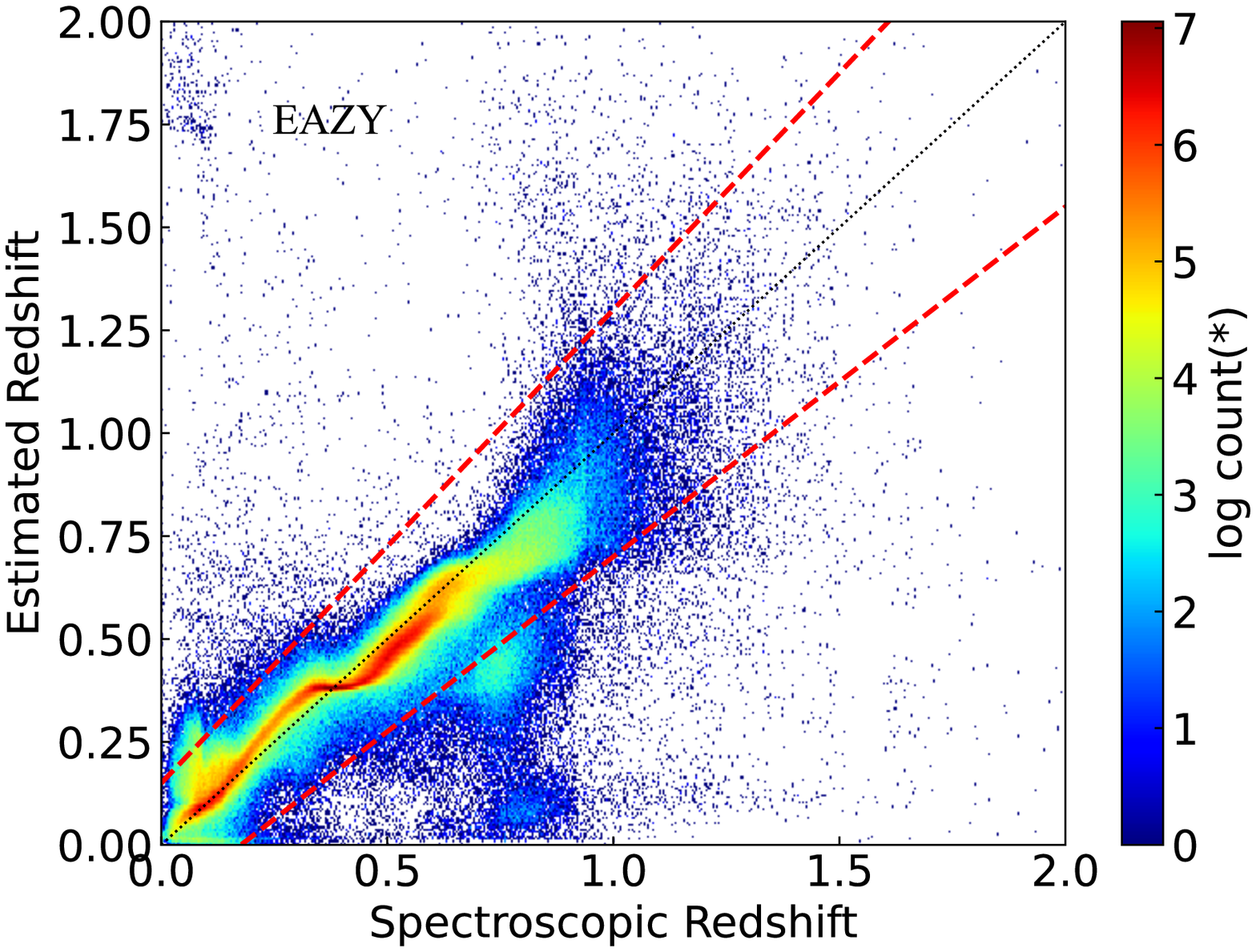}             		
     			\includegraphics[height=5cm,width=7cm]{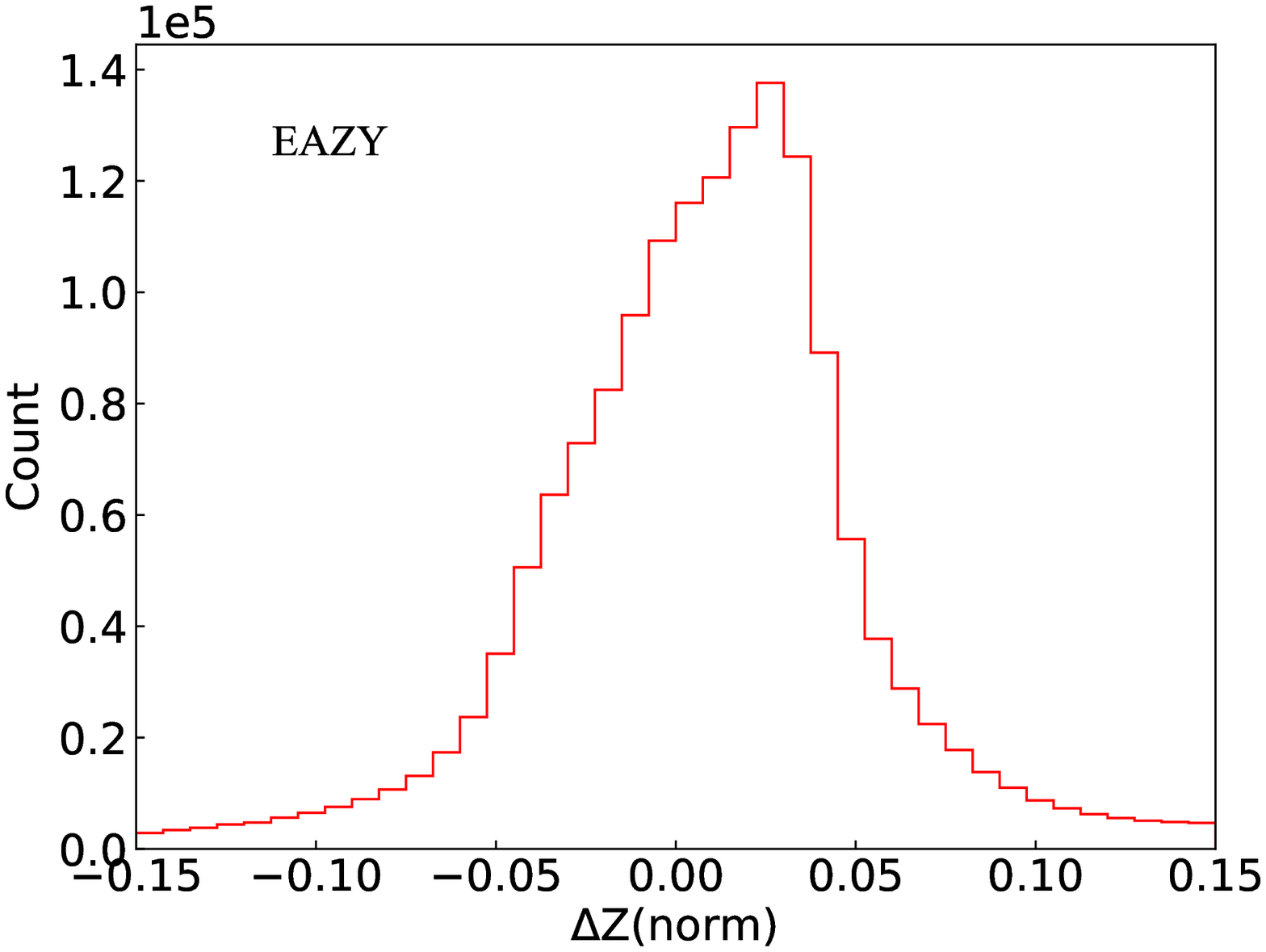}
     		}
     \subfigure[For the subsample DSW\_north]{
     	\includegraphics[height=5cm,width=7cm]{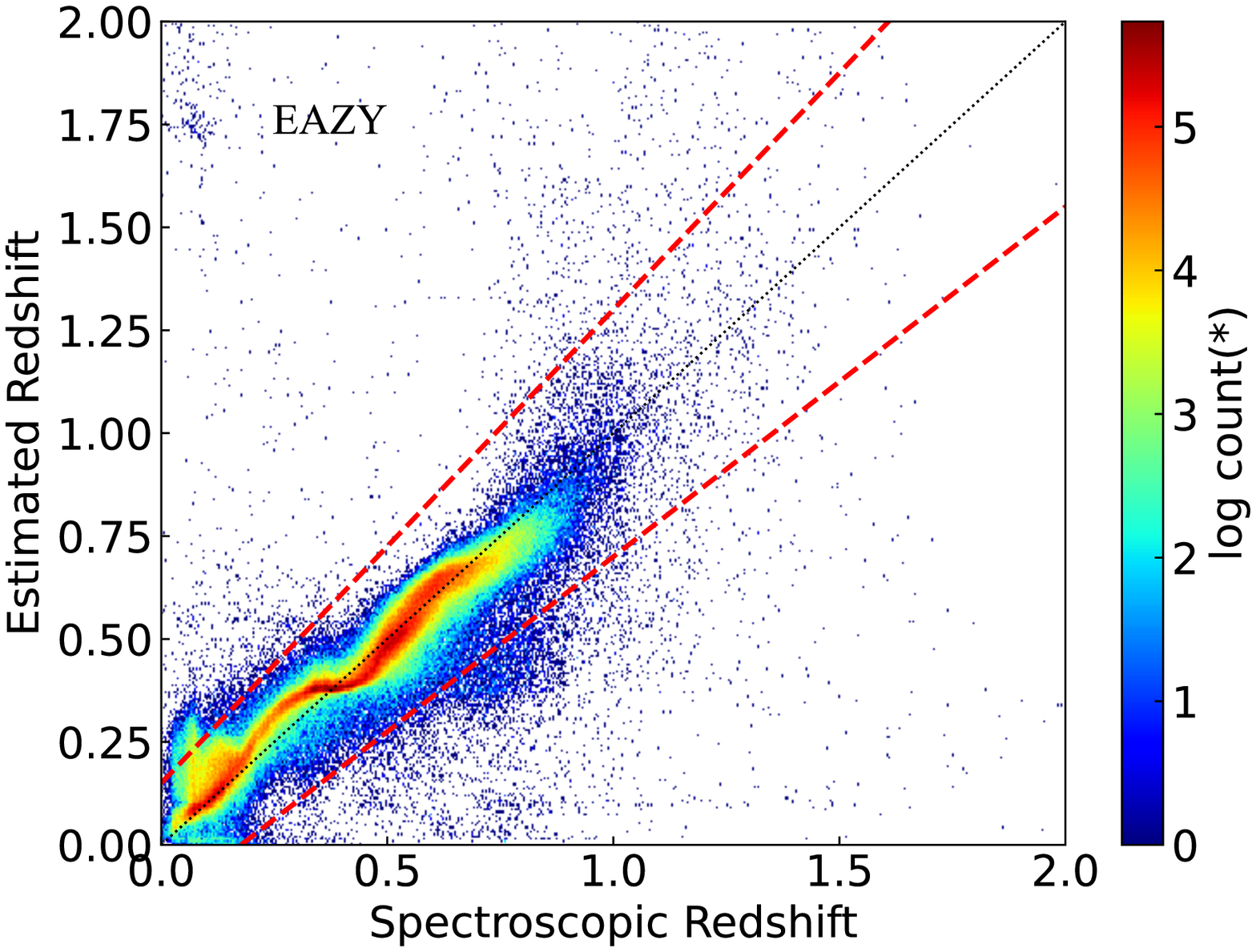}
     	\includegraphics[height=5cm,width=7cm]{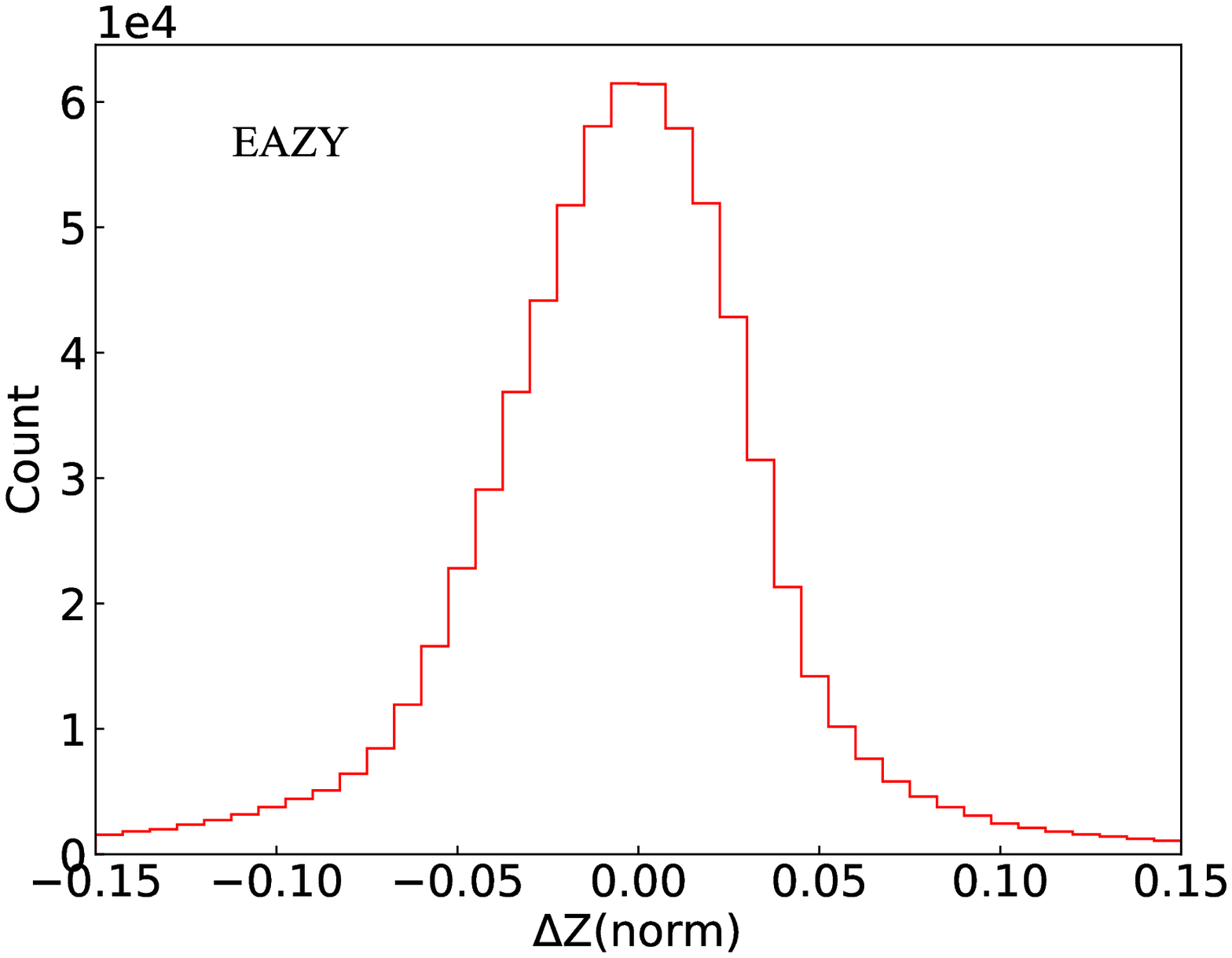}  	
     }
     	\caption{The scatter figure and $\Delta{\rm z(norm)}$ distribution of estimated photometric redshifts and spectroscopic redshifts for the subsamples of the DSW sample by {\small EAZY}. In the scatter figure, the red dashed line represents $\Delta \rm {z(norm)}=\pm 0.3$ separately. }
     	\label{fig2}
     \end{figure*}

\subsection{Photometric redshift estimation by {\small CATBOOST}}
Compared with template-fitting approach, the obvious characteristics of machine learning methods is that they need perform model training on known samples to establish a regressor for redshift estimation. The training process mainly includes feature selection, model parameter optimization and model validation.

\subsubsection{Feature selection}
In machine learning algorithms, feature selection is the key step influencing the performance of photometric redshift estimation.
According to the parameter description listed in Table~2, we choose features related to model magnitudes. These features include optical features $g$, $r$, $z$, $g-r$, $r-z$, $g-z$, and infrared related features $W1$, $W2$, $g-W1$, $r-W1$, $z-W1$, $g-W2$, $r-W2$, $z-W2$, $W1-W2$. Referring to the optimal feature selection procedure in \citet{Li2022}, we evaluate the importance score of each feature by {\small CATBOOST}, and sort them by the importance score. The importance score of each feature is shown in Figure~3. According to the rank of features, we select the top four features as initial input pattern to train, then add one feature in turn to the input pattern for training, and record all performance. $MSE$, $\sigma_{\rm NMAD}$, and $O$ with different input features is described in Figure~4. Figure~4 indicates that $MSE$, $\sigma_{\rm NMAD}$, and $O$ achieves the minimum when the number of input features is seven, and as the number of features increases, the performance fluctuates slightly. Therefore, the best input pattern (Pattern I, 7 features) is $g-r$, $g$, $r-z$, $r$, $r-W1$, $z-W1$, $W1-W2$, when only considering model magnitudes and colours.

\begin{figure}
\centering
\includegraphics[height=6cm,width=8cm]{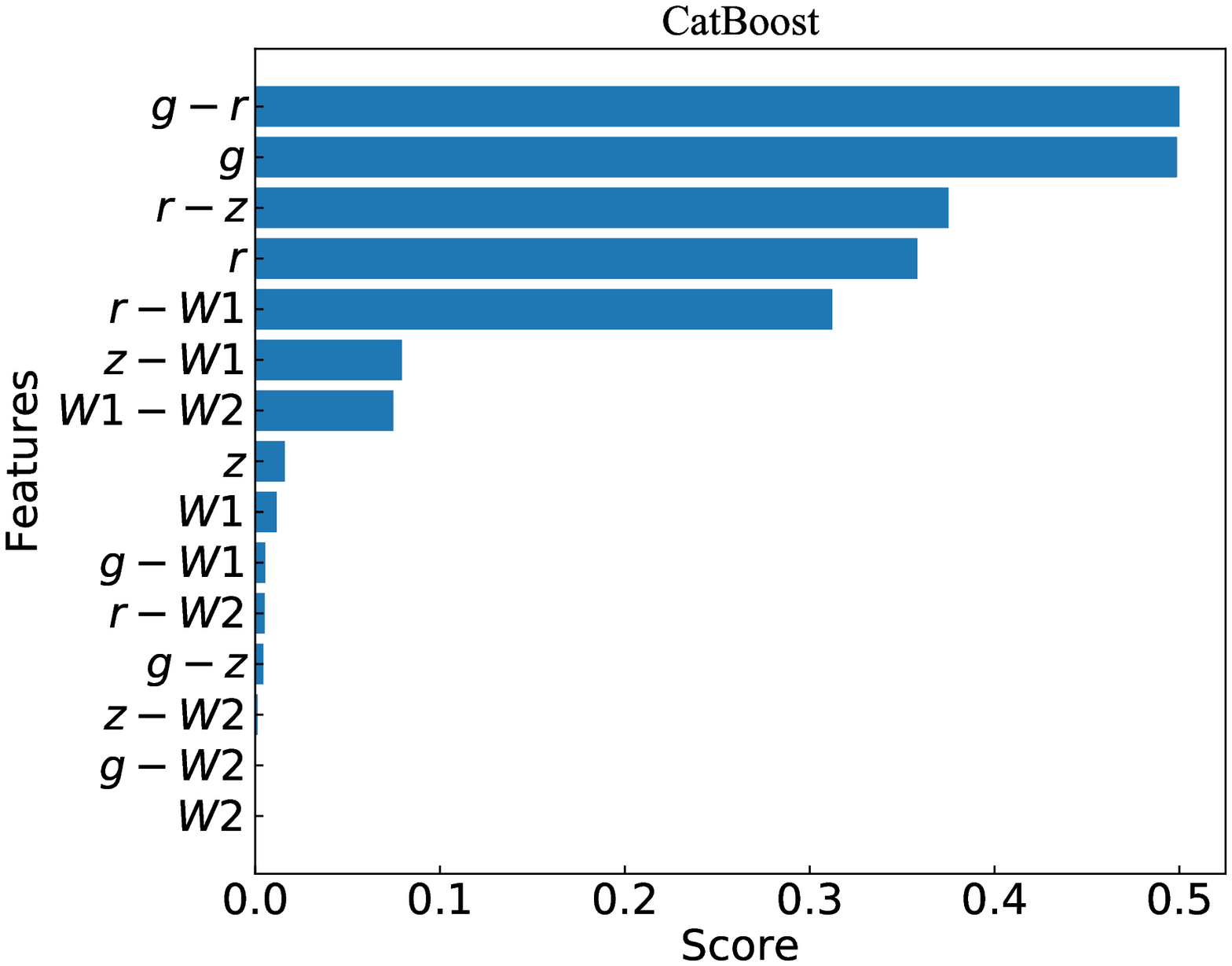}
\caption{The feature importance for the DSW sample by {\small CATBOOST}.}
\label{fig3}
\end{figure}

\begin{figure}
	\centering
	\includegraphics[height=12cm,width=8cm]{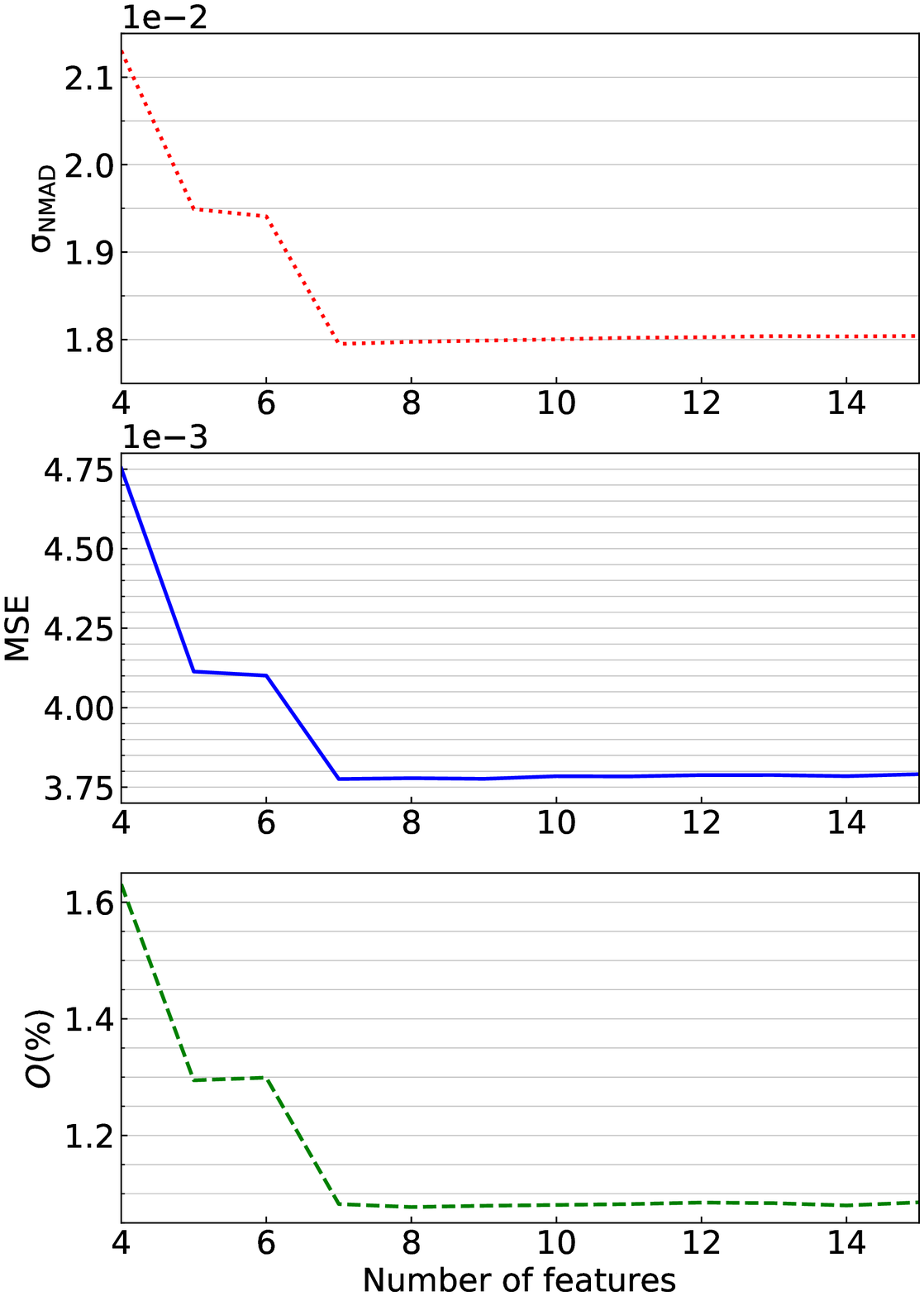}
	\caption{$\rm MSE$, $\sigma_{\rm NMAD}$, and $O$ as a function of different features by {\small CATBOOST}.}
	\label{fig4}
\end{figure}

Then, we mainly consider different aperture magnitudes as features. In the DESI DR9 catalogue, columns $apflux\_\{g,r,z\}$ contain eight different aperture fluxes in apertures of radius [0.5, 0.75, 1.0, 1.5, 2.0, 3.5, 5.0, 7.0] arc sec, and columns $apflux\_w1$, $apflux\_w2$ contain five different aperture fluxes in apertures of radius [3, 5, 7, 9, 11] arc sec. We convert these fluxes to AB magnitudes. The similar feature importance procedure as the above is applied.  The selected features contain $g-ap\_g\_1$, $r-ap\_r\_1$, $z-ap\_z\_1$, $W1-ap\_W1\_1$, $W2-ap\_W2\_1$
and the difference between two adjacent aperture magnitudes (e.g.
$ap\_g\_1-ap\_g\_2$, $ap\_g\_2-ap\_g\_3$, $ap\_g\_3-ap\_g\_4$, $ap\_g\_4-ap\_g\_5$, $ap\_g\_5-ap\_g\_6$, $ap\_g\_6-ap\_g\_7$ $ap\_g\_7-ap\_g\_8$) and the colors for different apertures in two adjacent bands (e.g. $ap\_g\_1-ap\_r\_1$, $ap\_r\_1-ap\_z\_1$, $ap\_z\_1-ap\_W1\_1$, $ap\_W1\_1-ap\_W2\_1$), which are ranked according to the importance score and regarded as Pattern II (60 features). If Pattern II is utilized, the best performance is $MSE=0.0056$, $\sigma_{\rm NMAD}=0.0250$. It is not better than the optimal model features (Pattern I).

Finally, combining the above obtained features related to model and aperture magnitudes, we obtain 75 features in total. We train again to get the importance score of each feature by {\small CATBOOST} and sort them. Those features with the score value of 0 are excluded. When 53 features are kept and defined as Pattern III, we get the best performance with $MSE=0.0034$ and $\sigma_{\rm NMAD}=0.0173$ for this case. Table~5 shows that aperture information is helpful to improve the accuracy of photometric redshift estimation of galaxies. About the detailed definition of Patterns I, II and III refers to the Appendix.

Because galaxies are extended objects, they have different sizes based on their luminosity and distance, they usually seems bigger when they are bright and near. For a special galaxy, it has definitive size which corresponds to appropriate aperture, and aperture features influence physical parameter measurement of galaxies. Therefore we also consider aperture features as input for photo-$z$ estimation. Different types of galaxies have different shapes of spectral energy distribution (SED), and the SED shape of a galaxy will move to longer wavelength by a factor $1+z$ at redshift $z$. The position of the obvious features (e.g. Lyman and Balmer breaks) as well as strong emission and absorption lines in a galaxy directly affects the shape of its low-resolution spectra, which in turn has effect on the photo-$z$ estimation. The flux becomes larger where emission line appears, while the flux turns smaller where absorption line emerges. For our sample, the redshift range is from 0 to 2. Taking Balmer breaks for example, the distribution of Balmer breaks is from 4000$\AA$ to 12000$\AA$, including $g$, $r$ and $z$ band. As shown in Figure~1, most of our sample occupy at low redshift, so the features related to $g$ and $r$ show higher score, while those related to $W2$ obtain lower score.

 \begin{table*}
 	\begin{center}
 		\caption[]{The performance of photometric redshift estimation by {\small CATBOOST} with default model parameters.\label{tab:fsel}}
 		\begin{tabular}{rccccccccc}
 			\hline
 			Sample &   Pattern   &  MSE &MAE& $\mathrm{Bias}$ & ${\sigma}_{\rm NMAD}$& ${\sigma}_{{\Delta}\rm z(norm)}$&${\delta}_{0.3}$(\%)&$O$(\%)&Time(s)\\
 			\hline
 			DSW   &   Pattern I   & 0.0038&0.0304&$1.0 \times 10^{-6}$ &0.0180&0.0406&99.72&1.08&88 \\
 			DSW   &	  Pattern II  & 0.0056 &0.0412&$2.85 \times 10^{-4}$ &0.0250&0.0504&99.64&2&154 \\ 	
 			DSW   &	  Pattern III  &0.0034 &0.0291&$3.6 \times 10{^4}$ &0.0173&0.0384&99.76&0.95 &143 \\ 		
 			\hline
 				\multicolumn{10}{l}{\tiny{$^a$ Pattern I represents $g - r$, $g$, $r - z$, $r$, $r - W1$, $z - W1$, $W1 - W2$ (7 features).}} \\
 				\multicolumn{10}{l}{\tiny{$^b$ Pattern II (60 features) and Pattern III (53 features) are described in the APPENDIX.}} \\
 				
 		\end{tabular}
 	\end{center}
 \end{table*}

\subsubsection{Model parameter optimization}
Model parameter optimization is a very complex and important task. {\small CATBOOST} is easy to achieve good performance with default parameters. Hence, we only choose the maximum depth of individual tree ($depth$) and the maximum number of trees ($iterations$) for optimization, for other parameters, the default values are adopted. For each training, we adopt 5-fold cross validation to obtain average $MSE$, $MAE$, $Bias$, ${\sigma}_{\rm NMAD}$, ${\sigma}_{{\Delta}\rm {z(norm)}}$, ${\delta}_{0.3}$, $O$ and running time. There are two steps to find optimal $depth$ and $iterations$. Firstly, we set $iterations$ as 1000 (default value), and in turn $depth$ is from 3 to 15. We separately perform this model parameter optimization procedure on the input patterns I and III. $MSE$, ${\sigma}_{\rm NMAD}$ and $O$ of each training with different $depth$ is displayed in Figure~5. Figure~5 shows that the performance with Pattern I reaches the best with $MSE=0.0037$ and $O=1.07$ per~cent when $depth=8$, the performance with Pattern III reaches the optimum with $MSE=0.0034$ and $O=0.92$  per~cent when $depth=12$. Secondly, we set $iterations$ in [1000, 2000, 3000, 4000, 5000] in turn. For Pattern I, the optimal $iterations$ is 4000 when $depth=8$. For Pattern III, the optimal $iterations$ is 5000 when $depth=12$. The performance metrics with optimal model parameters are listed in Table~6. Table~6 suggests that the performance with Pattern III is superior to that with Pattern I.

\begin{figure}
	\centering
	\includegraphics[height=12cm,width=8cm]{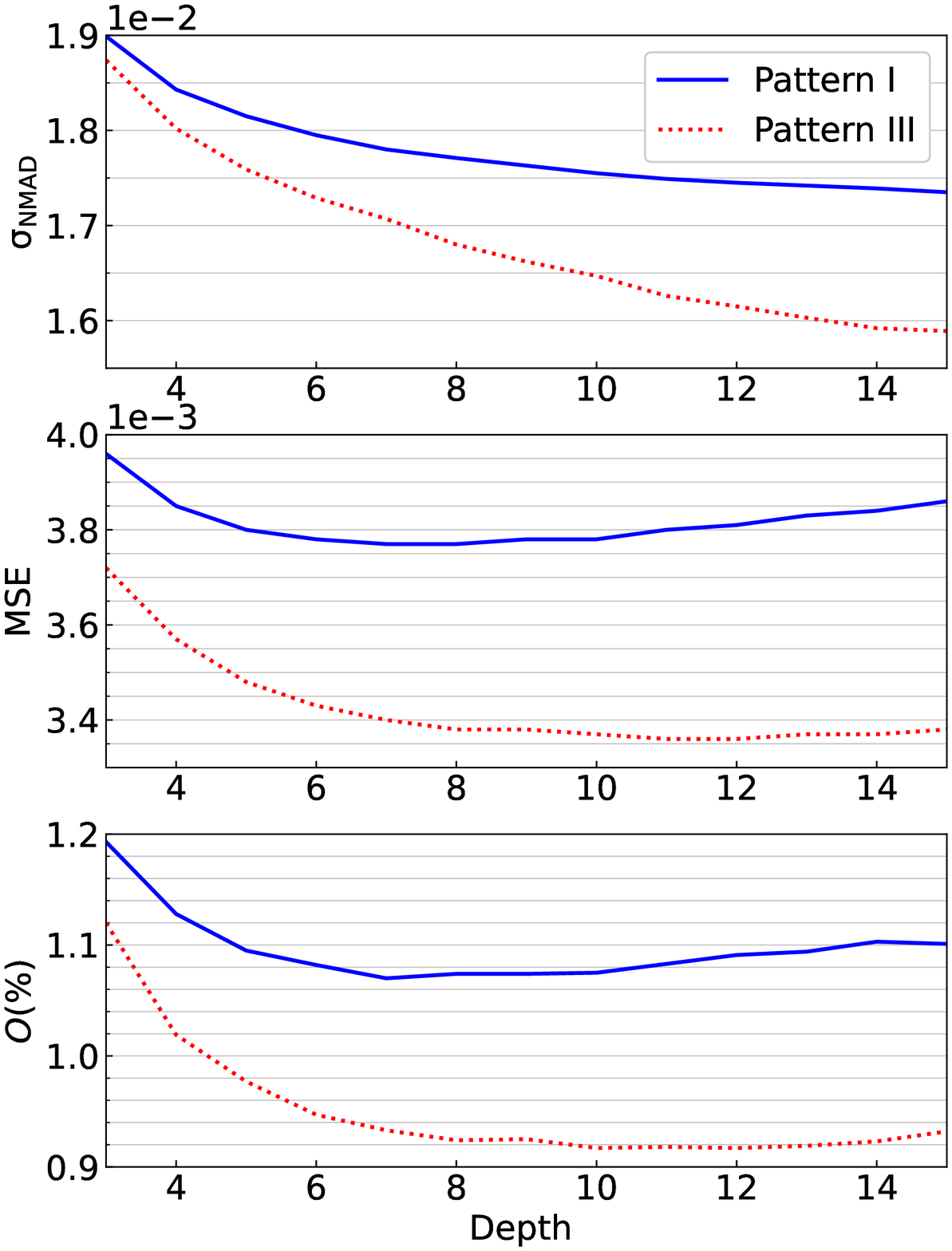}
	\caption{$\rm MSE$, ${\sigma}_{\rm NMAD}$ and $\rm O$ by {\small CATBOOST} with different $depth$. }
		\label{fig5}
	\end{figure}

       \begin{table*}
       	\small
       	\begin{center}
       		\caption[]{The performance of photometric redshift estimation with the best features and optimal model parameters by {\small CATBOOST}. \label{tab:confusion}}
       		\begin{tabular}{p{9mm}<{\centering}p{11mm}<{\centering}p{22mm}<{\centering}p{7mm}<{\centering}p{7mm}<{\centering}p{15mm}<{\centering}p{7mm}<{\centering}p{10mm}<{\centering}p{10mm}<{\centering}p{7mm}<{\centering}p{6mm}}
       			\hline      			
       			Sample & Pattern   & Model parameter&MSE & MAE& $\mathrm{Bias}$ & ${\sigma}_{\rm NMAD}$& ${\sigma}_{\rm{{\Delta}z(norm)} }$ &${\delta}_{0.3}$(\%)&$O$(\%)&Time(s)\\
       			\hline
       			DSW    & \text{Pattern I}  & $depth=8$ &0.0037 &0.0299&$4.8 \times10^{-5}$&0.0175 &0.0405&99.72&1.06&381 \\
       			& &  $iterations=4000$ &&&&&&&&\\
       			
     			DSW    & \text{Pattern III}  &  $depth=12$& 0.0032 &0.0272&$3.3\times10^{-4}$&0.0156 &0.0371&99.76&0.88&3484 \\
       			& & $iterations=5000$ &&&&&&&&\\   	         					     			
       			\hline
       			
       		\end{tabular}
       	\end{center}
       \end{table*}
%Last, with optimal model parameters, we train again for the sample DSW and evaluation itself, The scatter figure and  $\Delta{\rm z(norm)}$ distribution of estimated photometric redshifts and spectroscopic redshifts is shown in Figure~4.

Finally, we use the optimal pattern (Pattern III) and optimal model parameters ($depth=12$ and $iterations=5000$) of {\small CATBOOST} to train a regressor with the total sample, then use the same sample to evaluate. The performance is up to $MSE=0.0009$, $MAE=0.0182$, ${\sigma}_{\rm NMAD}=0.0128$, ${\sigma}_{\Delta\rm {z(norm)}}=0.0199$, $\delta_{0.3}=99.98$ per~cent, and $O=0.15$ per~cent. The scatter figure and $\Delta \rm {z(norm)}$ distribution of estimated photometric redshifts and spectroscopic redshifts are shown in Figure~6. Figure~6 implies that {\small CATBOOST} is reliable and efficient to estimate photometric redshifts of galaxies.

\begin{figure}
	\centering
	\includegraphics[height=6cm,width=7cm]{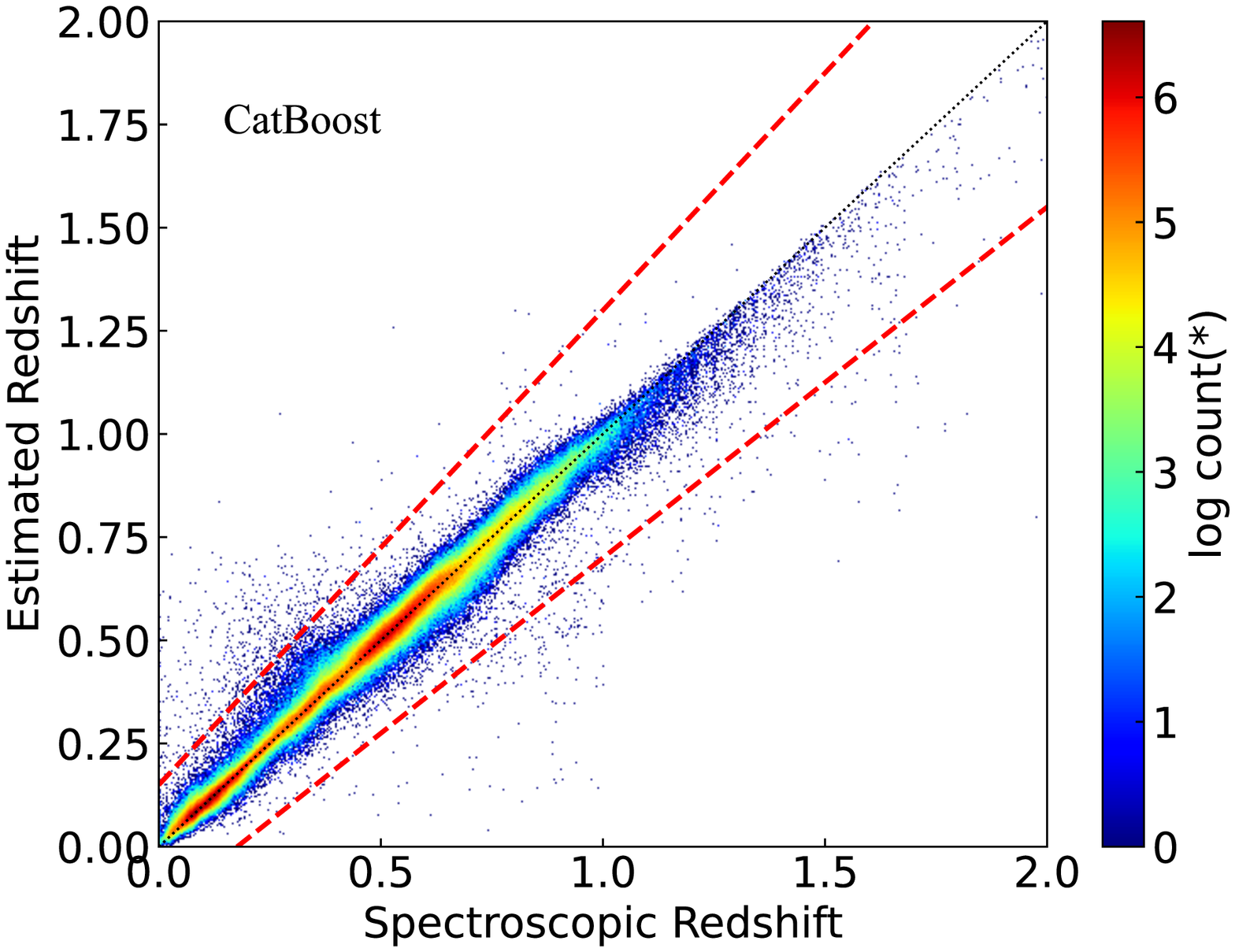}
	\includegraphics[height=6cm, width=7cm]{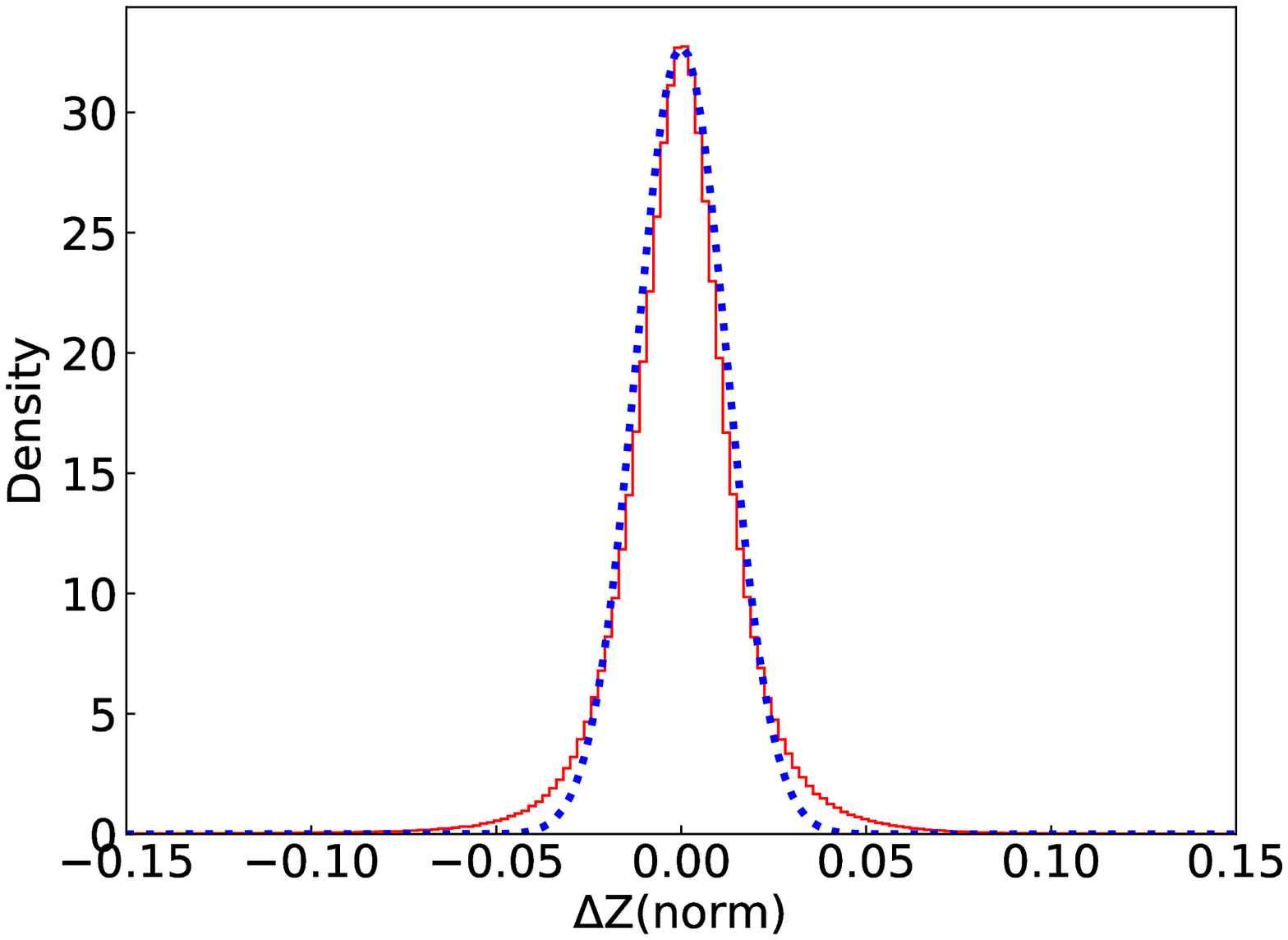}
	\caption{The evaluation performance of photometric redshift estimation with {\small CATBOOST}. In the scatter figure of photometric redshifts vs. spectroscopic redshifts, the red dashed line represents $\Delta \rm z(norm)=\pm0.3 $ separately. The distribution of $\Delta \rm {z(norm)}$ is between -0.15 and 0.15, the blue dotted curve is the best-fitting normal distribution.}
	\label{fig6}
\end{figure}

In order to assess the performance with magnitude and redshift for {\small CATBOOST}, Figure~7 depicts the metrics ($MSE$, $\sigma_{\rm NMAD}$, $O$) as a function of $r$ magnitude and redshift for the whole DSW sample, respectively. It shows that the three metrics become larger with the increase of magnitude, in other words, the performance of photometric redshift estimation declines when the celestial objects get fainter. It also indicates that the three metrics wave with different redshifts, and rise when redshift is larger than about 0.6, that is to say, the performance of the regressor is influenced by redshifts and decreases at higher redshift.

\begin{figure*}
	\centering
	\includegraphics[height=12cm,width=8cm]{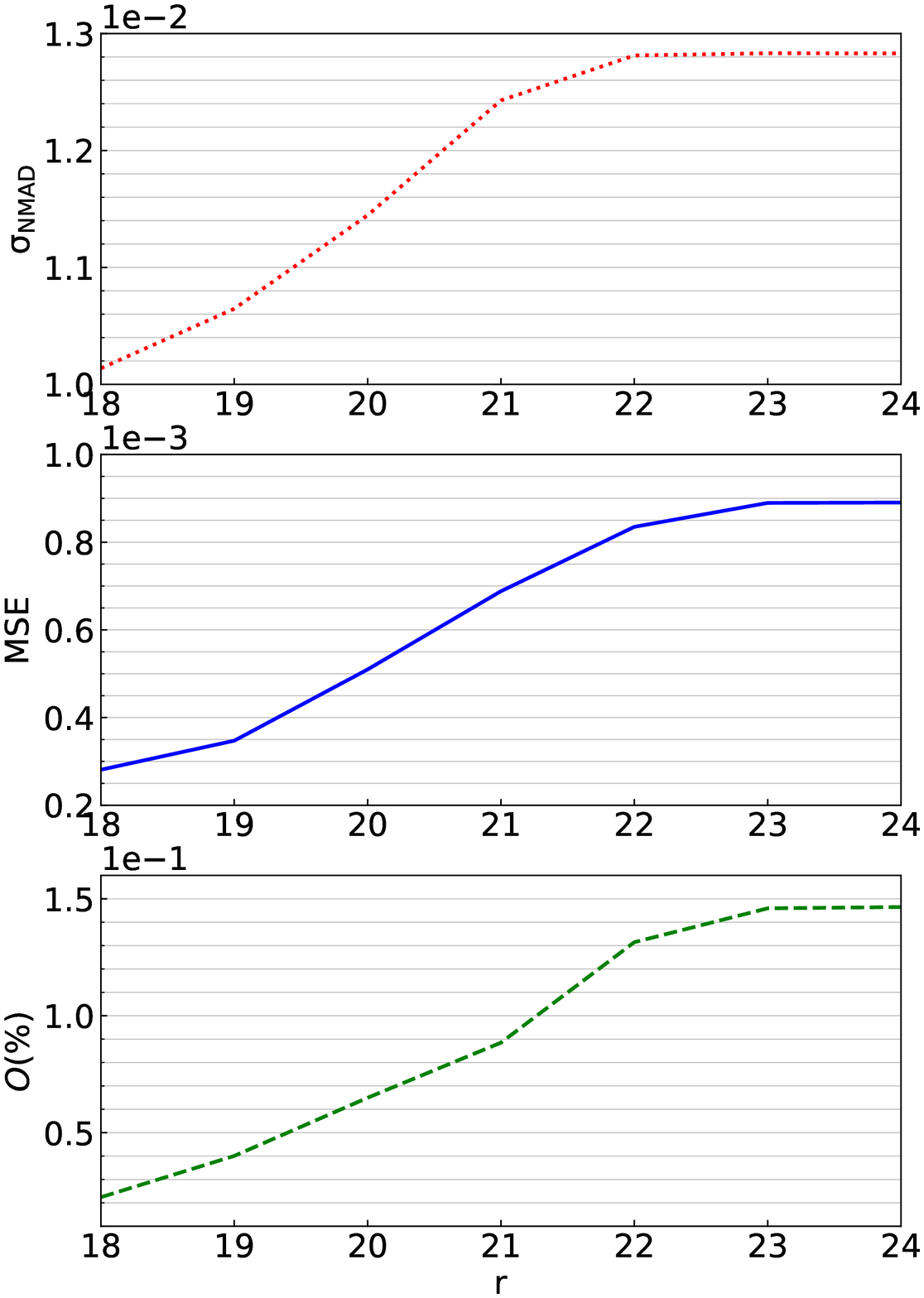}
	\includegraphics[height=12cm, width=8cm]{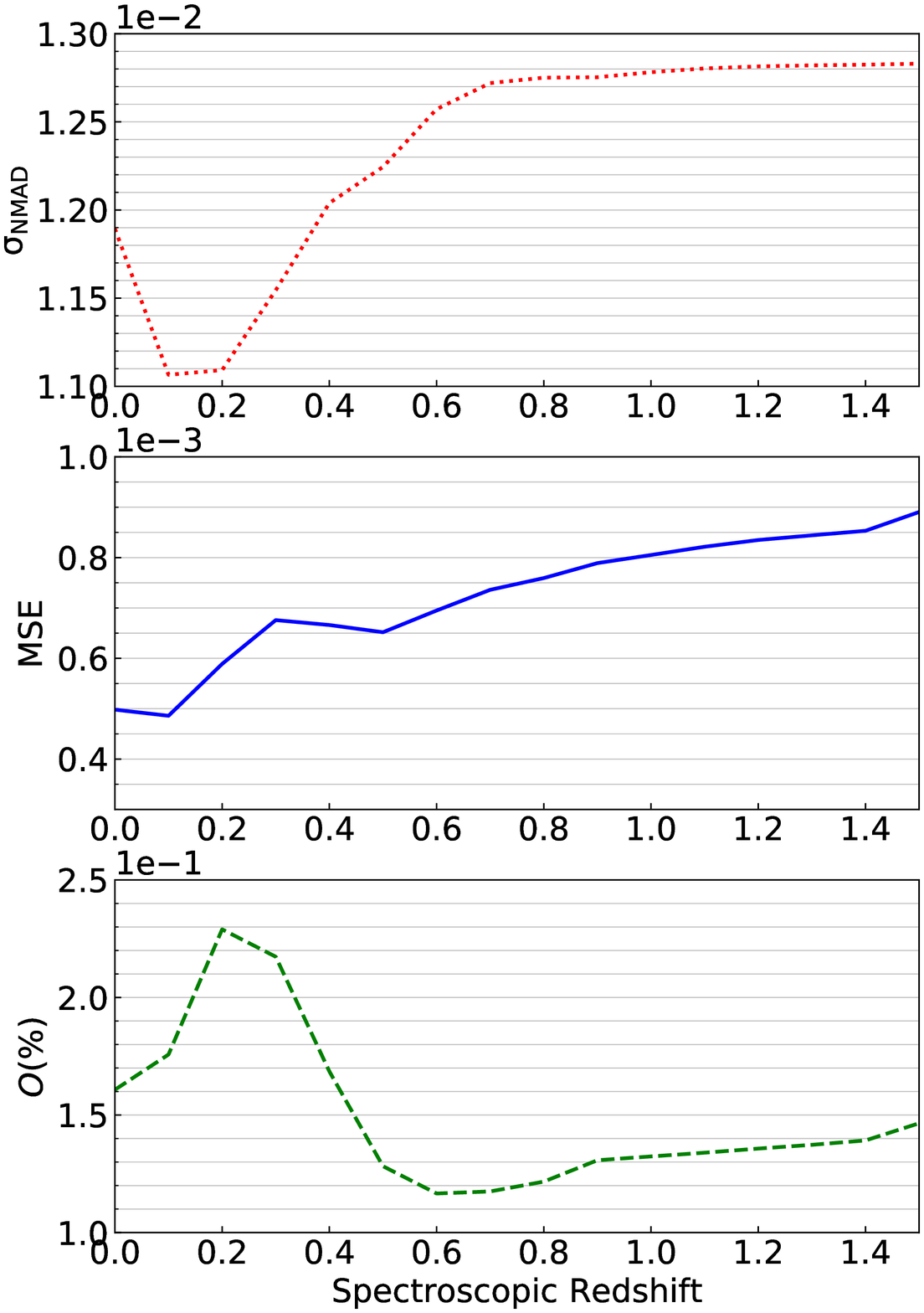}
	\caption{Left panel: the $\sigma_{\rm NMAD}$, $MSE$, $O$ as a function of $r$ magnitude for the whole DSW sample. Right panel: the $\sigma_{\rm NMAD}$, $MSE$, $O$ as a function of  spectroscopic redshifts for the whole DSW sample.}
	\label{fig7}
\end{figure*}

\subsubsection{Model Test}
After the {\small CATBOOST} regressor is established, we use three external test samples (S\_LAMOST, S\_GAMA, S\_WiggleZ) to test the performance of this regressor. The performance metrics for the test samples are described in Table~7. The scatter of photometric redshifts vs. spectroscopic redshifts are displayed respectively in Figure~8. Table~7 and Figure~8 show the estimation accuracy of photometric redshifts for different test samples is different, obviously the prediction accuracy with the S\_LAMOST and S\_GAMA samples are better than that with the S\_WiggleZ sample. It further proves that the prediction accuracy is affected by the magnitude and redshifts of samples. The sources with brighter magnitude and lower redshift have better prediction accuracy, which is consistent with the fact that most of the S\_LAMOST and S\_GAMA samples are brighter and at lower redshifts while most of the S\_WiggleZ sample are fainter. For the S\_WiggleZ sample, the performance as a function of magnitude and redshift is shown in Figure~9.  Comparing Figure~9 to Figure~7, the trends with magnitude are almost similar but the trend with magnitude in Figure~9 is higher than in Figure~7; the trends with redshift are clearly different; when redshift is larger than 0.6 the trends with redshift both become gradual. This result may be due to the selection effect that the S\_WiggleZ sample is inclined to emission galaxies.

      \begin{table*}
      	\begin{center}
      		\caption[]{The performance of the external test samples.  \label{tab:confusion}}
      		\begin{tabular}{rcccccccccc}
      			\hline
      			Test Sample &  &  MSE & MAE& $\mathrm{Bias}$ & ${\sigma}_{\rm NMAD}$& ${\sigma}_{{\Delta}\rm z(norm)}$&${\delta}_{0.3}$(\%)&$O$(\%)\\
      			\hline      		
	      		S\_LAMOST &  & 0.0010 & 0.0142 & 0.0022 & 0.0113 & 0.0242 & 99.86 & 0.49 \\
	      		S\_GAMA & & 0.0022 & 0.0273 & 0.0015 & 0.0202 & 0.0351 & 99.62 & 1.06 \\
	      		S\_WiggleZ & & 0.0208 & 0.0981 & -0.031 & 0.0669 & 0.0661 & 99.03 & 9.04 \\
      			\hline
      		\end{tabular}
      	\end{center}
      \end{table*}

\begin{figure*}
	\centering
      	\subfigure[For the LAMOST sample. ]{
      		\includegraphics[height=6.1cm]{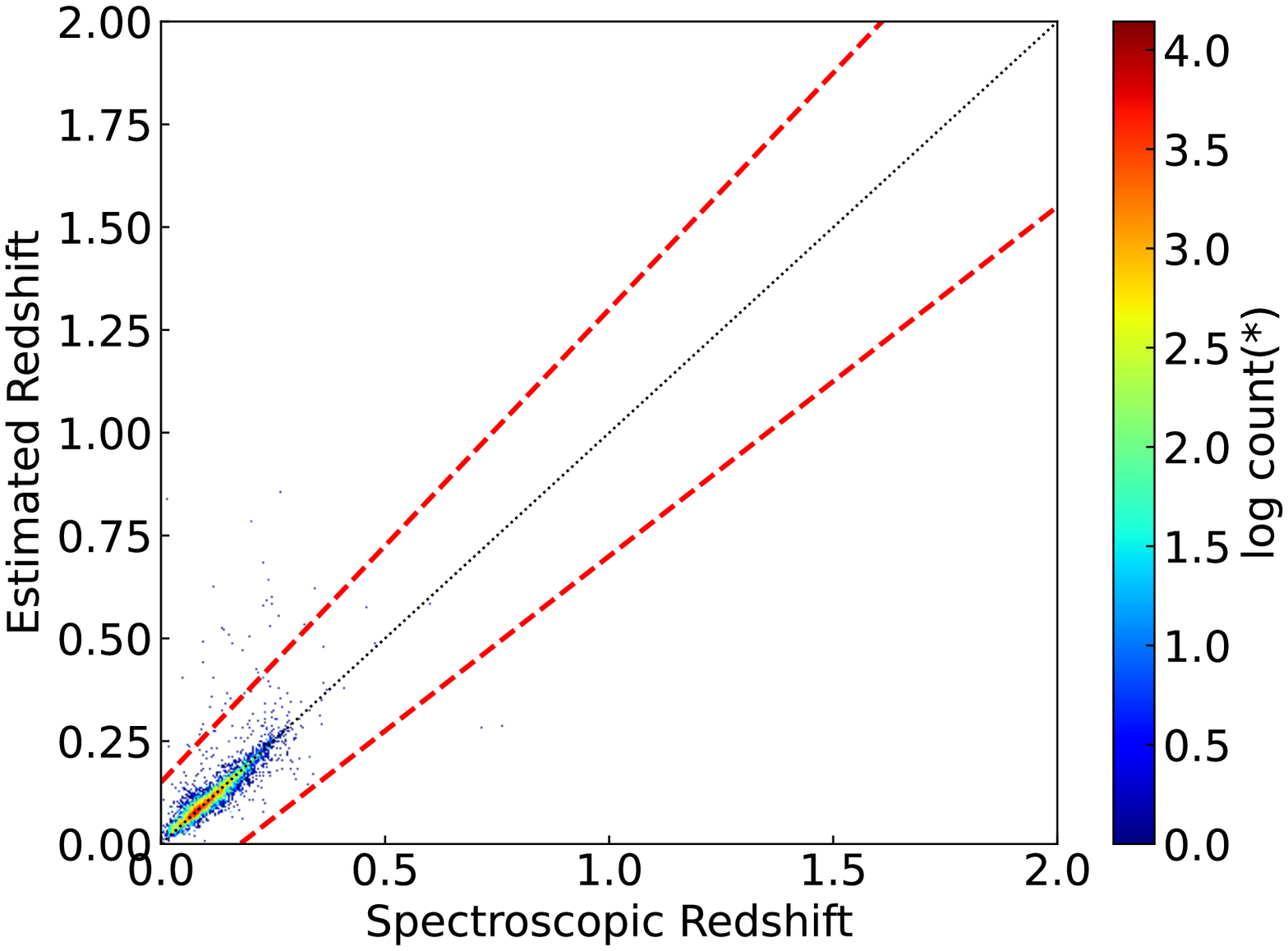}
      		\includegraphics[height=6.5cm, width=7cm]{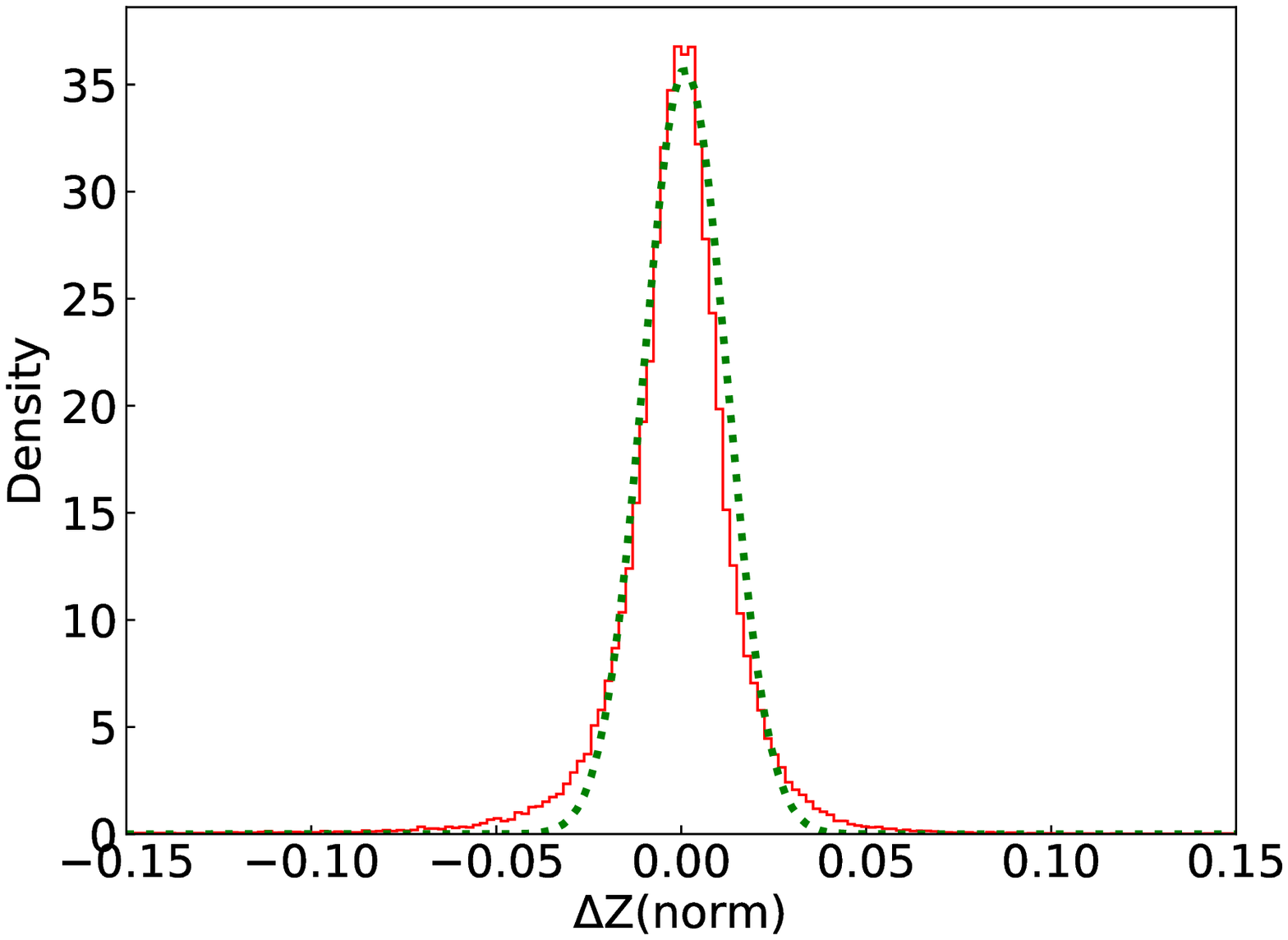}
      	}
      	
      		\subfigure[For the GAMA sample. ]{
      			\includegraphics[height=6.1cm]{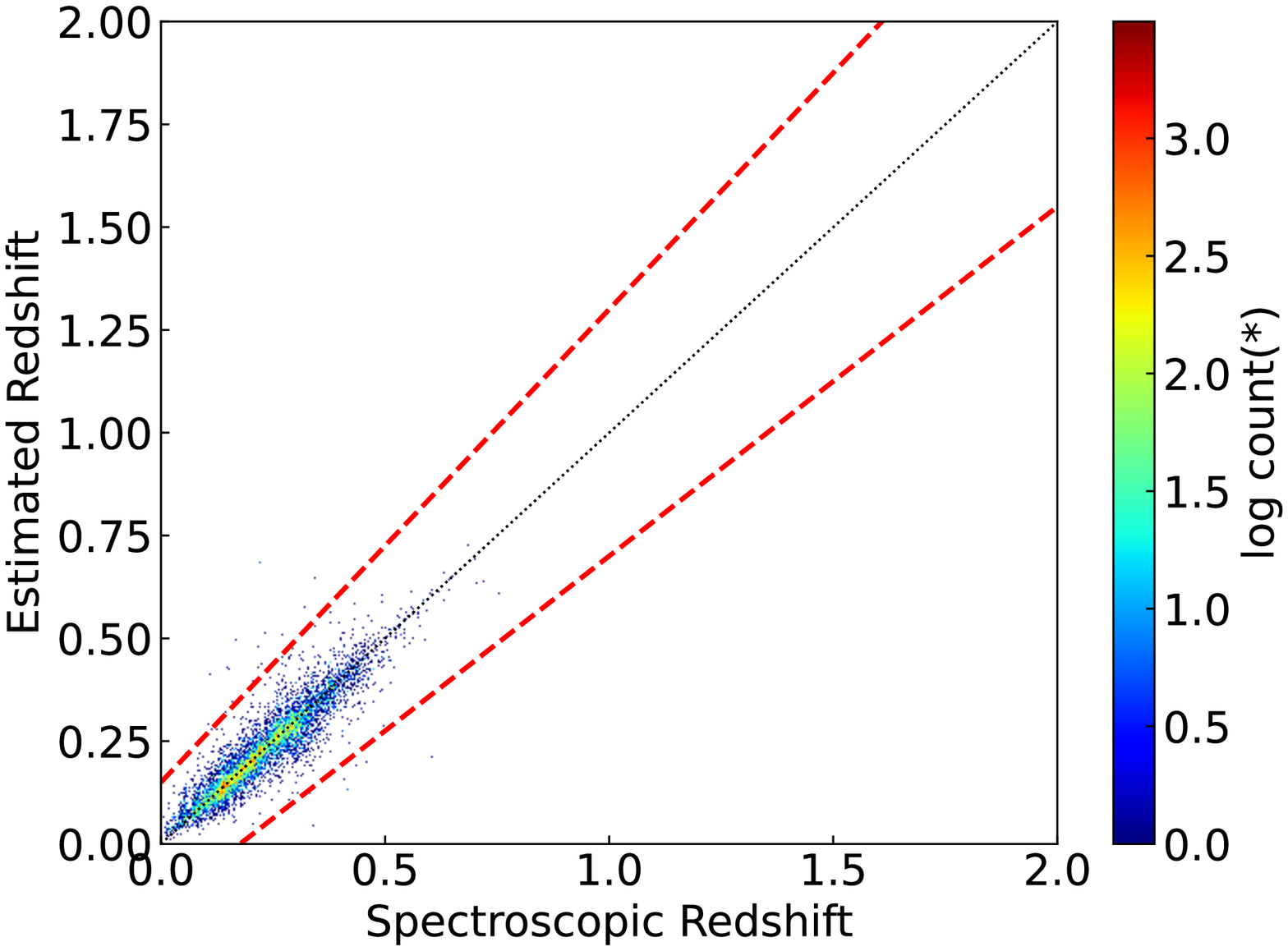}
      			\includegraphics[height=6.5cm,width=7cm]{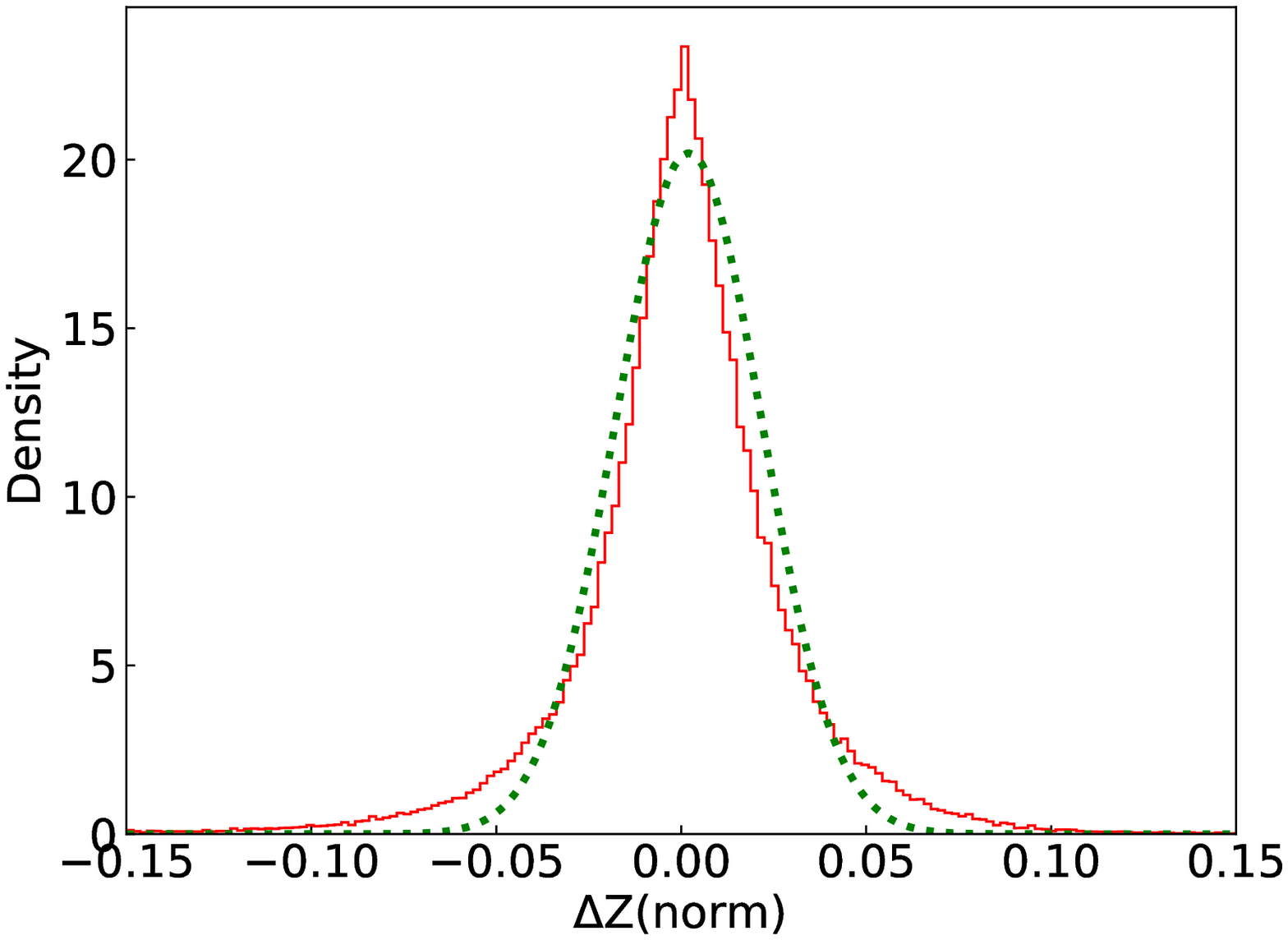}
      		}
      		
      			\subfigure[For the WiggleZ sample. ]{
      				\includegraphics[height=6.1cm]{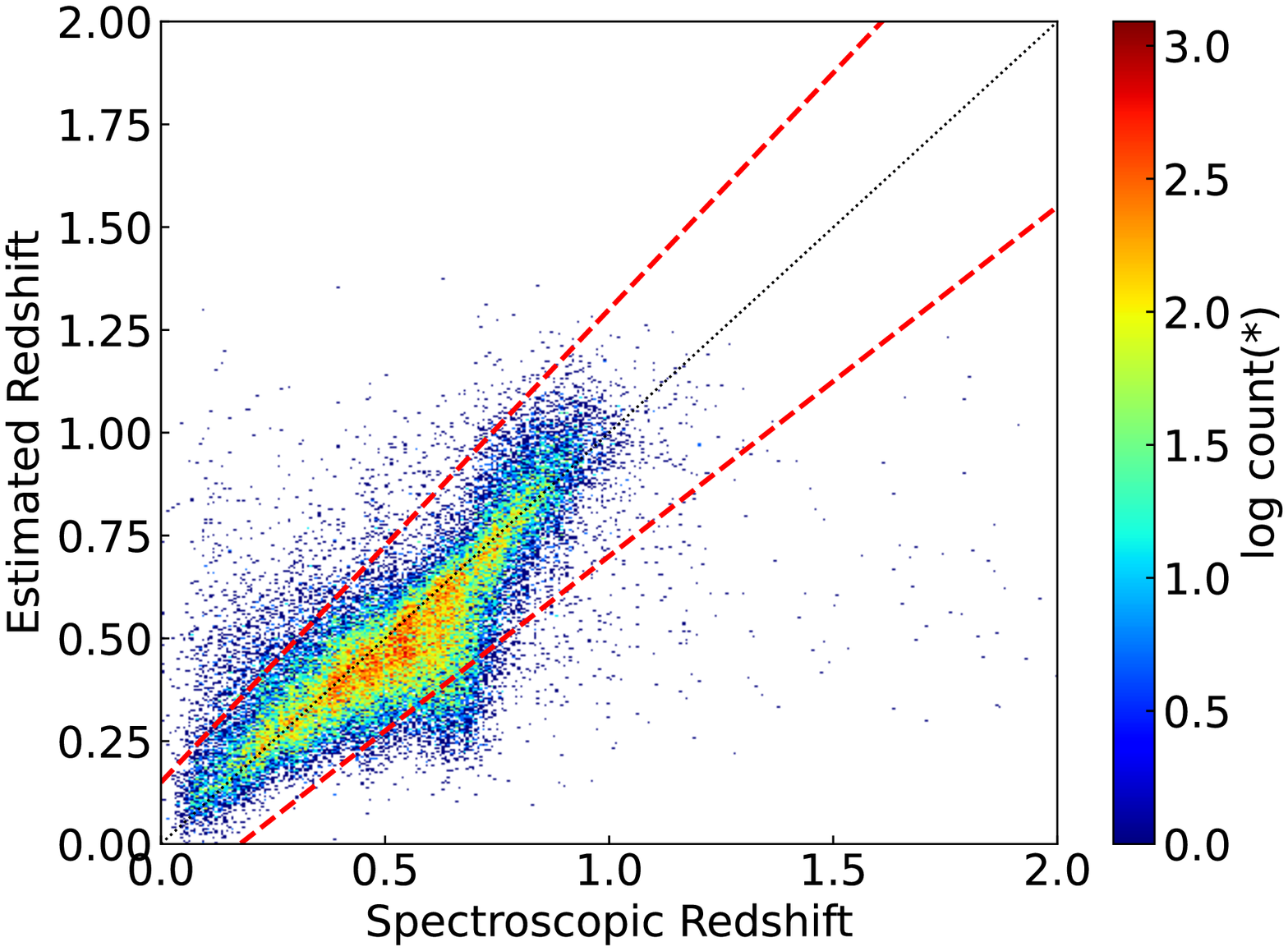}
      				\includegraphics[height=6.5cm,width=7cm]{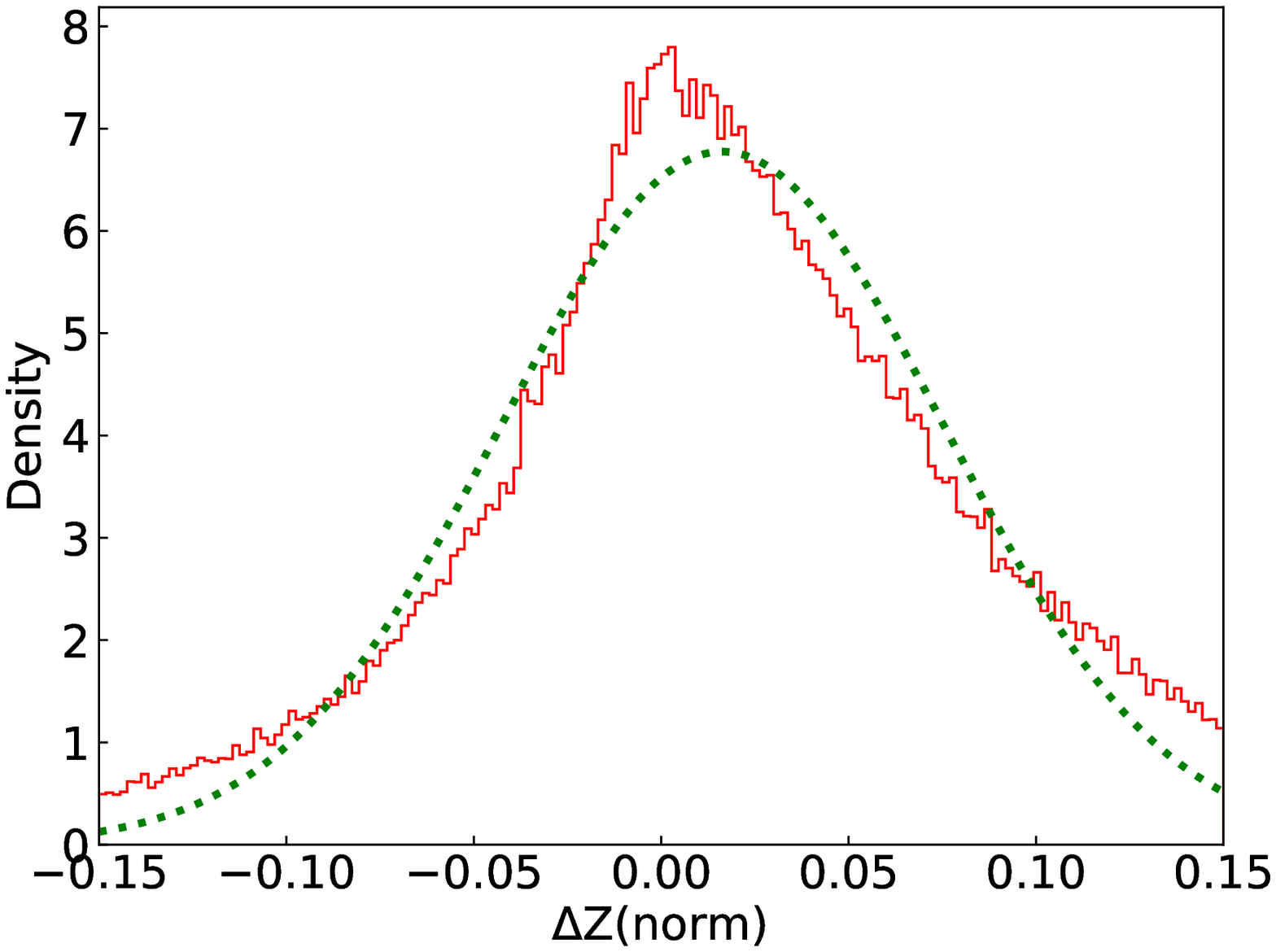}
      			}

\caption{Photometric redshift estimation with the external test samples S\_LAMOST, S\_GAMA, S\_WiggleZ.
In the scatter figure of photometric redshifts vs. spectroscopic redshifts, the red dashed line
represents $\Delta \rm {z(norm)}=\pm 0.3 $ separately. For the $\Delta \rm {z(norm)}$ distribution, red line for the $\Delta \rm {z(norm)}$ distribution and green dotted line is the best-fitting normal distribution.}
      	\label{fig8}
      \end{figure*}

\begin{figure*}
	\centering
	\includegraphics[height=12cm,width=8cm]{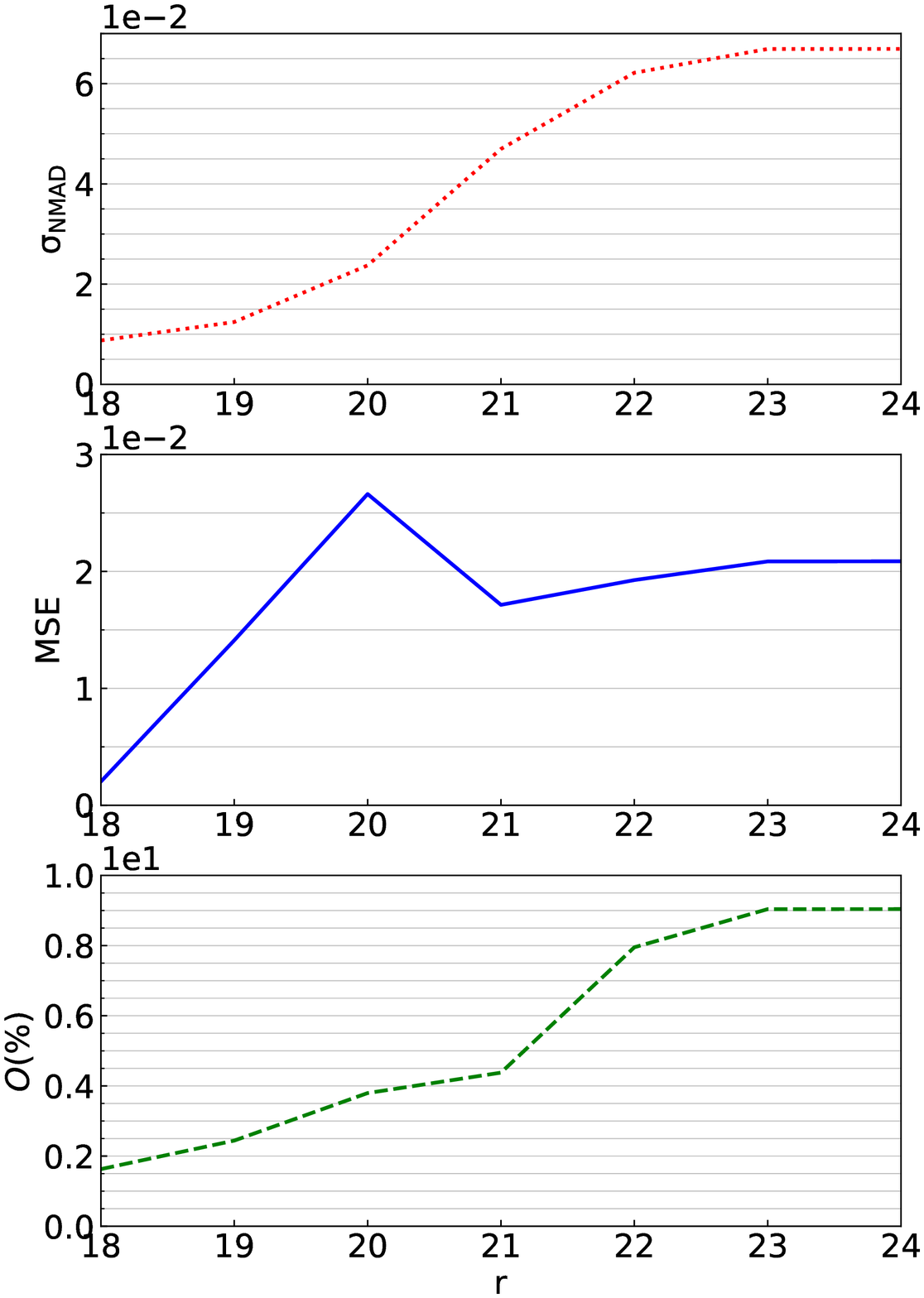}
	\includegraphics[height=12cm, width=8cm]{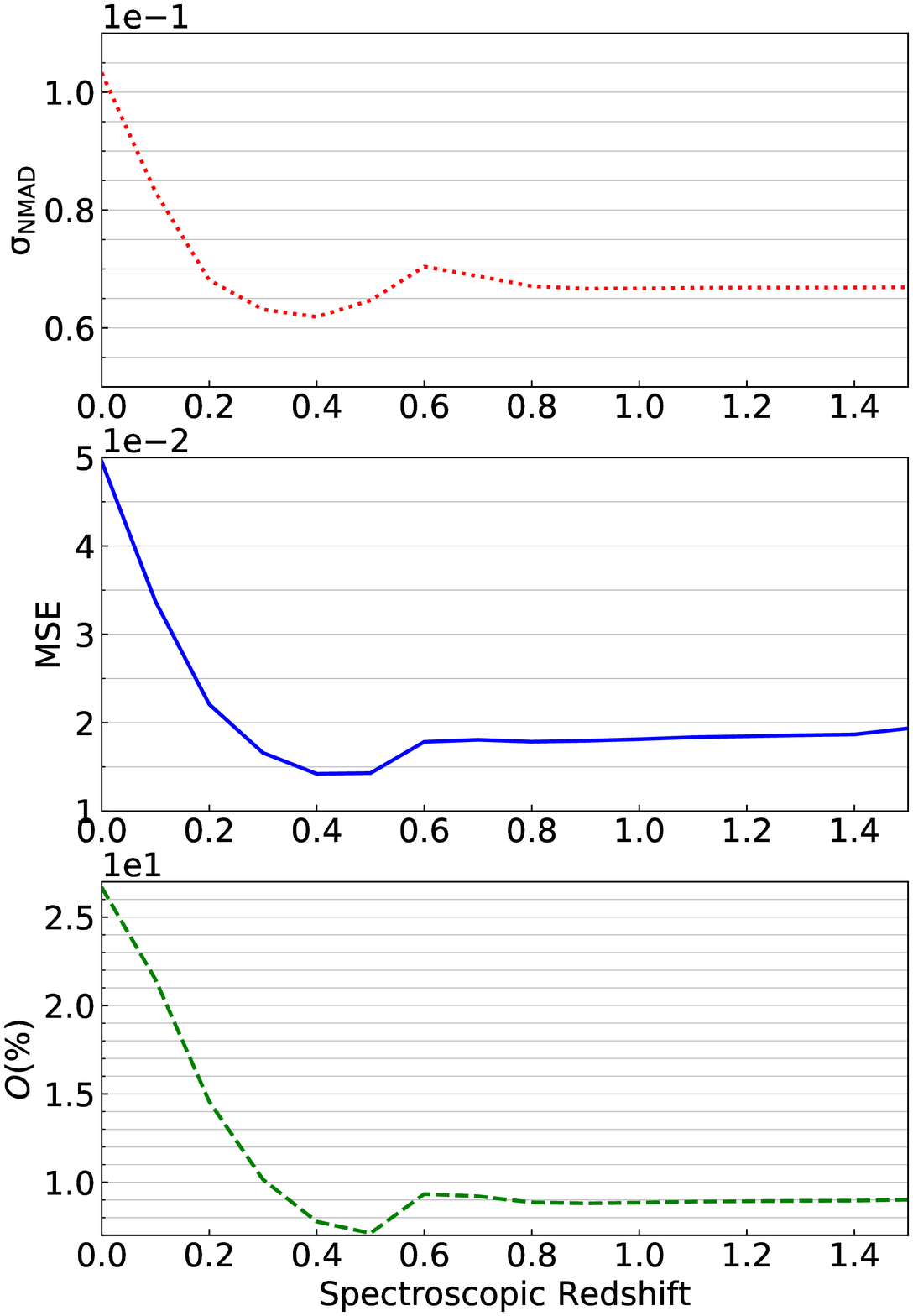}
	\caption{Left panel: the $\sigma_{\rm NMAD}$, $MSE$, $O$ as a function of $r$ magnitude for the WiggleZ sample. Right panel: the $\sigma_{\rm NMAD}$, $MSE$, $O$ as a function of  spectroscopic redshifts for the WiggleZ sample.}
	\label{fig9}
\end{figure*}

\subsection{Comparison and discussion}
Although {\small CATBOOST} is optimal compared to XGBoost and Random Forest (RF) in \citet{Li2022}, the performance is closely related to used samples. We adopt multi-layer perceptron (MLP) and RF to train regressors again with Pattern III respectively. The python code packages of MLP and RF are provided by scikit-learn (\citealt{scikit-learn}). Table~8 shows their optimal performance. Comparing the results of different methods in Table~8, it is found that the performance of {\small CATBOOST} is superior to that of MLP in respect of all metrics, {\small CATBOOST} outperforms RF except Bias, moreover {\small CATBOOST} has much shorter training time than RF.

	      \begin{table*}
	      	\small
	      	\begin{center}
	      		\caption[]{The performance of photometric redshift estimation with the best features and optimal model parameters by different methods. \label{tab:confusion}}
	      		\begin{tabular}{p{6mm}<{\centering}p{9mm}<{\centering}p{12mm}<{\centering}p{27mm}<{\centering}p{5mm}<{\centering}p{5mm}<{\centering}p{15mm}<{\centering}p{7mm}<{\centering}p{10mm}<{\centering}p{10mm}<{\centering}p{7mm}<{\centering}p{6mm}<{\centering}}
	      			\hline      			
	      			Sample & Method &Pattern   & Model parameter&MSE & MAE& $\mathrm{Bias}$ & ${\sigma}_{\rm NMAD}$& ${\sigma}_{{\Delta}\rm z(norm)}$&${\delta}_{0.3}$(\%)&$O$(\%)&Time(s)\\
	      			\hline	      			
	      			DSW  &{\tiny CATBOOST}  & \text{Pattern III}  &  $depth=12$& 0.0032 &0.0272&$3.3\times10^{-4}$&0.0156 &0.0371&99.76&0.88&3484 \\
	      			& && $iterations=5000$ &&&&&&&&\\
	      			DSW &MLP    & \text{Pattern III}  &  $hidden\_layer\_sizes=100$ & 0.0044 &0.0367&$1.1\times10^{-3}$&0.0246 &0.0439&99.71&1.38&464 \\
	      			& & &$max\_iter=200$ &&&&&&&&\\   		         					
	      			%DSW & RF & \text{Pattern III} & max\_depth=None & 0.0034 & 0.0279 & $4.8 \times 10^{-4}$ & 0.0158 & 0.0383 & 99.74 & 0.96 & 18910 \\
		      		%& & &n\_estimators=300 & &&&&&&&\\
	      			DSW & RF & \text{Pattern III} & $max\_depth=12$ & 0.0037 & 0.0306 & $1.9 \times 10^{-4}$ & 0.0183 & 0.0401 & 99.74 & 1.04 & 34044 \\
	      			& & &$n\_estimators=300$ & &&&&&&&\\
	      			\hline
	      			
	      		\end{tabular}
	      	\end{center}
	      \end{table*}
		
\citet{zhou2021} explored the clustering of DESI-like luminous red galaxies by photometric redshifts and adopted Random Forest method to estimate photometric redshifts of DESI DR8. Comparing our work with \citet{zhou2021}, the differences are methods and data, as shown in Table~9. Tables~8-9 both further suggests that {\small CATBOOST} outperforms Random Forest for our case. \citet{zhou2021} presented photometric redshifts for DESI DR8 with Random Forest while we provide the photometric redshift measurement for DESI DR9 by means of {\small CATBOOST} and {\small EAZY}.

\begin{table*}
\small
\begin{center}
\caption[]{Comparison of photometric redshift estimation in our work with that in \citet{zhou2021}. \label{tab:confusion}}
		  	    		\begin{tabular}{p{24mm}<{\centering}p{50mm}<{\centering}p{15mm}<{\centering}p{1mm}<{\centering}p{1mm}<{\centering}p{15mm}<{\centering}p{1mm}<{\centering}p{10mm}<{\centering}p{10mm}<{\centering}}
		  	    			\hline      			
			  	    	Work	 &Data  &Method & &  & ${\sigma}_{\rm NMAD}$& &$O(0.10)$(\%)&\\
		  	    			\hline	      			
		  	    		    \citet{zhou2021}&DESI DR8 South ($MAG\_Z \le 21$)  &RF& &&0.0133 &&1.51&\\
									  	    		   &DESI DR8  North ($MAG\_Z \le 21$) &RF & &&0.0136 &&0.92&\\
									  	    		   &DESI DR8  South ($MAG\_Z \textgreater 21$) &RF & &&0.0725 &&24.6&\\
		  	    		
		  	    			Our work & DESI DR9 South ($MAG\_Z \le 21$) &{\small CATBOOST} &&&0.0126&&0.34&\\	    		         					
		  	    			  &DESI DR9 North ($MAG\_Z \le 21$ ) &{\small CATBOOST} &&&0.0130 & & 0.22&\\
		  	    			    &DESI DR9 South ($MAG\_Z \textgreater  21$) &{\small CATBOOST} &&&0.0148 & & 1.24&\\
		  	    			      &DESI DR9 North ($MAG\_Z \textgreater  21$) &{\small CATBOOST} &&&0.0132 & & 1.39&\\
		  	    			
		  	    			\hline	
		  	    				\multicolumn{8}{l}{\tiny{$^a$ $O(0.10)$ represents outlier fraction of $\left| \Delta{\rm z(norm)} \right| \textgreater 0.10$. } } \\  			
		  	    		\end{tabular}
		  	    	\end{center}
		  	    \end{table*}

\subsection{Application}
Based on the above experimental results for the known samples, we use {\small EAZY} and {\small CATBOOST} to predict photometric redshifts of galaxies from DESI DR9. We select all galaxy sources whose morphological classification types are equal to `REX', `EXP', `DEV' or `SER'. The number of sources amounts to 1 160 568 989. We estimate photometric redshifts of these galaxies using {\small EAZY} and {\small CATBOOST} with optimized parameters, respectively. {\small EAZY} can predict redshifts of galaxies ranging from 0 to 6. Table~10 lists the number of predicted DESI DR9 galaxies with $maskbits=0$ and $Q_{\rm z}<1$ in different redshift ranges by {\small EAZY}. As shown in Table~10, the number of galaxies with predicted redshifts above 2 occupies about 6 per cent. For predicted DESI DR9 galaxies with $maskbits=0$, $\rm g \le 24.0$, $\rm r \le 23.4$ and $\rm z \le 22.5$, the predicted results by {\small CATBOOST} are more reliable when estimated redshifts is below 2, those by {\small EAZY} are more confident when estimated redshifts is greater than 2 and $Q_{\rm z}<1$. As a result, {\small EAZY} is very useful for finding high redshift galaxies. While for  the out-of-range sources with $g > 24.0$, $r > 23.4$ or $z > 22.5$, the predicted redshifts are only for reference. The whole predicted results by {\small EAZY} and {\small CATBOOST} are saved in 1107 files, which is of great value for the further study on the characteristics and evolution of galaxies. The result file link address is https://doi.org/10.12149/101162. For simplicity, Table~11 only lists 20 rows of predicted results, but all column of fluxes and magnitudes related are saved in the result files. 

\begin{table*}
	\begin{center}
		\caption[]{The number of predicted DESI DR9 galaxies with $maskbits=0$ and $Q_{\rm z}<1$ in different redshift ranges by {\small EAZY}.}
		\begin{tabular}{lllccc}
			\hline
			Method      & $0<$redshift $<2$ &$2\le$ redshift$< 3.5$&$3.5\le$redshift$<4.5$     &$4.5\le$redshift$<5.5$ &redshift$\ge 5.5$ \\
			\hline
			% {\small EAZY}&380 235 202&18 201 953 &3 747 786 &1 788 811 &144 278 \\
			 {\small EAZY} &348 853 174&15 826 023&3 328 406& 1 635 656 & 127 971 \\
				\hline
		\end{tabular}
	\end{center}
\end{table*}

\begin{table*}
	\begin{center}
		\caption[]{The estimated photometric redshifts of DESI DR9 galaxies, $z\_cb$ is predicted redshift by {\small CATBOOST}, $z\_eazy$ is predicted redshift by {\small EAZY}, $Q_{\rm z}$ demonstrates photometric redshift estimation quality of {\small EAZY}, nfilt is the number of used filters.}
		\begin{tabular}{lllcccccc}
			\hline
			release & brickid & objid & RA & Dec & z\_cb   & z\_eazy & $Q_{\rm z}$ & $nfilt$  \\
			\hline
			9010 &465328 &3933 &140.60547174695245 &23.89775480146321 &0.447 &0.277&5.420&5 \\
			9010 &465328 &3935 &140.60570892573824 &23.913236038787858 &0.980 &1.395&8.351&5 \\
			9010 &465328 &3936 &140.60576126216455 &24.08768860870403 &0.520 &0.375&0.047&4 \\
			9010 &465328 &3937 &140.60576728408122 &24.075655633141213 &0.949 &1.206&3.747&5 \\
			9010 &465328 &3938 &140.60577110445337 &24.033382424270663 &0.871 &1.142&4.221&4 \\
			9010 &465328 &3940 &140.60592807141936 &23.998381611411066 &0.650 &1.125&56.022&4 \\
			9010 &465328 &3942 &140.60615861124148 &23.94613561864017 &0.969 &1.677&13.574&5 \\
			9010 &465328 &3944 &140.60620883781942 &23.912058728862267 &0.889 &0.932&2.866&5 \\
			9010 &465328 &3948 &140.60638882279335 &24.028406631122056 &1.140 &1.305&1.543&5 \\
			9010 &465328 &3949 &140.60641713623545 &23.87420843970829 &0.913 &1.804&0.011&5 \\
			9010 &465328 &3954 &140.60659960507638 &23.910351365034238 &0.844 &0.500&0.333&5 \\
			9010 &465328 &3955 &140.60663973978004 &24.007290234431604 &0.858 &0.850&1.715&5 \\
			9010 &465328 &3956 &140.60683707462877 &24.04563942636601 &0.738 &0.711&1.145&4 \\
			9010 &465328 &3957 &140.6068563283938 &24.110469165376028 &0.824 &0.884&1.728&5 \\
			9010 &465328 &3958 &140.60685843145077 &24.106992421044172 &0.468 &2.298&3.353&4 \\
			9010 &465328 &3960 &140.6068938439817 &24.117431904938048 &0.503 &1.766&39.261&4 \\
			9010 &465328 &3964 &140.60719039910862 &24.04818401589685 &0.605 &0.320&0.889&4 \\
			9010 &465328 &3965 &140.60724459654958 &24.057913639613552 &0.953 &1.741&0.882&4 \\
			9010 &465328 &3967 &140.6073643677021 &24.03165350182948 &0.246 &0.297&5.305&4 \\
			9010 &465328 &3968 &140.60753251569648 &24.066669913889974 &0.483 &2.658&10.795&5 \\
			
			\hline
		\end{tabular}
	\end{center}
\end{table*}

%	    \begin{table*}
%	    	\small
%	    	\begin{center}
%	    		\caption[]{The performance of photometric redshift estimation. \label{tab:confusion}}
%	    		\begin{tabular}{p{24mm}<{\centering}p{20mm}<{\centering}p{10mm}<{\centering}p{10mm}<{\centering}p{20mm}<{\centering}p{15mm}<{\centering}p{7mm}<{\centering}p{10mm}<{\centering}p{10mm}<{\centering}}
%	    			\hline      			
%	    			Sample &predicted redshift  &MSE & MAE& $\mathrm{Bias}$ & ${\sigma}_{\rm NMAD}$& ${\sigma}_{{\Delta}\rm z(norm)}$&${\delta}_{0.3}$(\%)&$O$(\%)\\
%	    			\hline	      			
%	    		    TrueSample\\($MAG\_Z<21$)  &z\_cb&0.0013 &0.0210&$2.5\times10^{-4}$&0.0138 &0.0199&99.96&0.29\\
%	    			TrueSample\\($MAG\_Z<21$) &Z\_PHOT\_MEAN &0.0014 &0.0218&$-1.7\times10^{-4}$&0.0138 &0.0203&99.96&0.27\\
%	    			
%	    			TrueSample &z\_cb &0.0044 &0.0288&$-3.8\times10^{-3}$&0.0149 &0.0324&99.83&1.19\\	    		         					
%	    			TrueSample &Z\_PHOT\_MEAN&0.0037 &0.0291&$-4.2\times10^{-3}$&0.0152 &0.0312&99.83&0.92\\
%	    			
%	    			\hline	    			
%	    		\end{tabular}
%	    	\end{center}
%	    \end{table*}
%	

\section{Conclusions} \label{sec:conclusions}
We adopt template fitting and machine learning to predict photometric redshifts of galaxies. We discuss the advantages and disadvantages of both approaches. Template fitting methods are not affected by known samples and can predict galaxies with higher redshifts, while machine learning methods can achieve higher accuracy through model training, but due to the lack of high redshift galaxies in currently known samples, the high redshift galaxies will be predicted to be low redshift ones. For low redshift galaxies (redshift $z<2$ for our case), {\small CATBOOST} is a good choice to estimate their photometric redshifts; for high redshift ones, {\small EAZY} is more appropriate. Thus, we utilize the two methods to predict photometric redshifts of all galaxies from DESI Legacy Imaging Surveys DR9. The estimated photometric redshifts of galaxies will be very valuable for future research, and can help spectroscopic sky survey projects (e.g. SDSS, DESI) for follow up observation to spot more high redshift galaxies.

\section{Acknowledgements}
We are very grateful to the referee's constructive suggestions. This work is supported by National Natural Science Foundation of China (NSFC)(grant Nos. 11803055, 12273076, 12103070, 11573019, 11873066, 12133001, 11433005), the Joint Research Fund in Astronomy ( U1931132, U1531246, U1731125, U1731243, U1731109) under cooperative agreement between the NSFC and Chinese Academy of Sciences (CAS), the 14th Five-year Informatization Plan of Chinese Academy of Sciences (CAS-WX2021SF-0204) and the science research grants from the China Manned Space Project with Nos. CMS-CSST-2021-A04 and CMS-CSST-2021-A06. Data resources are supported by Chinese Astronomical Data Center (NADC), CAS Astronomical Data Center and Chinese Virtual Observatory (China-VO). This work is supported by Astronomical Big Data Joint Research Center, co-founded by National Astronomical Observatories, Chinese Academy of Sciences and Alibaba Cloud.

 The Guoshoujing Telescope (the Large Sky Area Multi-object Fiber Spectroscopic Telescope, LAMOST) is a National Major Scientific Project built by the Chinese Academy of Sciences. Funding for the project has been provided by the National Development and Reform Commission. LAMOST is operated and managed by the National Astronomical Observatories, Chinese Academy of Sciences.

The Legacy Surveys consist of three individual and complementary projects: the Dark Energy Camera Legacy Survey (DECaLS; Proposal ID \#2014B-0404; PIs: David Schlegel and Arjun Dey), the Beijing-Arizona Sky Survey (BASS; NOAO Prop. ID \#2015A-0801; PIs: Zhou Xu and Xiaohui Fan), and the Mayall z-band Legacy Survey (MzLS; Prop. ID \#2016A-0453; PI: Arjun Dey). DECaLS, BASS and MzLS together include data obtained, respectively, at the Blanco telescope, Cerro Tololo Inter-American Observatory, NSF's NOIRLab; the Bok telescope, Steward Observatory, University of Arizona; and the Mayall telescope, Kitt Peak National Observatory, NOIRR Lab. The Legacy Surveys project is honored to be permitted to conduct astronomical research on Iolkam Du'ag (Kitt Peak), a mountain with particular significance to the Tohono O'odham Nation.

NOIRLab is operated by the Association of Universities for Research in Astronomy (AURA) under a cooperative agreement with the National Science Foundation.

This project used data obtained with the Dark Energy Camera (DECam), which was constructed by the Dark Energy Survey (DES) collaboration. Funding for the DES Projects has been provided by the U.S. Department of Energy, the U.S. National Science Foundation, the Ministry of Science and Education of Spain, the Science and Technology Facilities Council of the United Kingdom, the Higher Education Funding Council for England, the National Center for Supercomputing Applications at the University of Illinois at Urbana-Champaign, the Kavli Institute of Cosmological Physics at the University of Chicago, Center for Cosmology and Astro-Particle Physics at the Ohio State University, the Mitchell Institute for Fundamental Physics and Astronomy at Texas A\&M University, Financiadora de Estudos e Projetos, Fundacao Carlos Chagas Filho de Amparo, Financiadora de Estudos e Projetos, Fundacao Carlos Chagas Filho de Amparo a Pesquisa do Estado do Rio de Janeiro, Conselho Nacional de Desenvolvimento Cientifico e Tecnologico and the Ministerio da Ciencia, Tecnologia e Inovacao, the Deutsche Forschungsgemeinschaft and the Collaborating Institutions in the Dark Energy Survey. The Collaborating Institutions are Argonne National Laboratory, the University of California at Santa Cruz, the University of Cambridge, Centro de Investigaciones Energeticas, Medioambientales y Techologicas-Madrid, the University of Chicago, University College London, the DES-Brazil Consortium, the University of Edinburgh, the Eidgenossische Technische Hochschule (ETH) Zurich, Fermi National Accelerator Laboratory, the University of Illinois at Urbana-Champaign, the Institut de Ciencies de l'Espai (IEEC/CSIC), the Institut de Fisica d'Altes Energies, Lawrence Berkeley National Laboratory, the Ludwig Maximilians Universitat Munchen and the associated Excellence Cluster Universe, the University of Michigan, NSF's NOIRLabb, the University of Nottingham, the Ohio State University, the University of Pennsylvania, the University of Portsmouth, SLAC National Accelerator Laboratory, Stanford University, the University of Sussex, and Texas A\&M University.

BASS is a key project of the Telescope Access Program (TAP), which has been funded by the National Astronomical Observatories of China, the Chinese Academy of Sciences (the Strategic Priority Research Program "The Emergence of Cosmological Structures" Grant \# XDB09000000), and the Special Fund for Astronomy from the Ministry of Finance. The BASS is also supported by the External Cooperation Program of Chinese Academy of Sciences (Grant \# 114A11KYSB20160057), and Chinese National Natural Science Foundation (Grant \# 11433005).

The Legacy Survey team makes use of data products from the Near-Earth Object Wide-field Infrared Survey Explorer (NEOWISE), which is a project of the Jet Propulsion Laboratory/California Institute of Technology. NEOWISE is funded by the National Aeronautics and Space Administration.

The Legacy Surveys imaging of the DESI footprint is supported by the Director, Office of Science, Office of High Energy Physics of the U.S. Department of Energy under Contract No. DE-AC02-05CH1123, by the National Energy Research Scientific Computing Center, a DOE Office of Science User Facility under the same contract; and by the U.S. National Science Foundation, Division of Astronomical Sciences under Contract No. AST-0950945 to NOAO.

We acknowledgment SDSS databases. Funding for the Sloan Digital Sky Survey IV has been provided by the Alfred P. Sloan Foundation, the U.S. Department of Energy Office of Science, and the Participating Institutions. SDSS-IV acknowledges support and resources from the Center for High-Performance Computing at the University of Utah. The SDSS web site is www.sdss.org. SDSS-IV is managed by the Astrophysical Research Consortium for the Participating Institutions of the SDSS Collaboration including the Brazilian Participation Group, the Carnegie Institution for Science, Carnegie Mellon University, the Chilean Participation Group, the French Participation Group, Harvard-Smithsonian Center for Astrophysics, Instituto de Astrof\'isica de Canarias, The Johns Hopkins University, Kavli Institute for the Physics and Mathematics of the Universe (IPMU) /University of Tokyo, Lawrence Berkeley National Laboratory, Leibniz Institut f\"ur Astrophysik Potsdam (AIP), Max-Planck-Institut f\"ur Astronomie (MPIA Heidelberg), Max-Planck-Institut f\"ur Astrophysik (MPA Garching), Max-Planck-Institut f\"ur Extraterrestrische Physik (MPE), National Astronomical Observatories of China, New Mexico State University, New York University, University of Notre Dame, Observat\'ario Nacional / MCTI, The Ohio State University, Pennsylvania State University, Shanghai Astronomical Observatory, United Kingdom Participation Group, Universidad Nacional Aut\'onoma de M\'exico, University of Arizona, University of Colorado Boulder, University of Oxford, University of Portsmouth, University of Utah, University of Virginia, University of Washington, University of Wisconsin, Vanderbilt University, and Yale University.

\section{Data availability}
The predicted photometric redshifts for DESI DR9 galaxies are saved in a repository and can be obtained by a unique identifier, part of which is indicated in Table~11. It is put in paperdata repository of NADC at http://paperdata.china-vo.org, and can be available with https://doi.org/10.12149/101162.

\section{Appendix: Pattern definition}
The patterns used in this paper is defined as follows:

Pattern I represents $g - r$, $g$, $r - z$, $r$, $r - W1$, $z - W1$, $W1 - W2$ (7 features).

Pattern II is
$ap\_r\_4-ap\_z\_4$, $ap\_g\_5-ap\_R\_5$, $z-ap\_z\_1$, $ap\_r\_6-ap\_z\_6$,  $g-ap\_g\_1$, $ap\_z\_4-ap\_z\_5$, $r-ap\_r\_1$, $ap\_z\_1-ap\_z\_2$,  $ap\_r\_1-ap\_r\_2$,  $ap\_r\_4-ap\_r\_5$, $ap\_g\_6-ap\_r\_6$, $ap\_r\_7-ap\_z\_7$, $ap\_z\_2-ap\_z\_3$, $ap\_r\_5-ap\_z\_5$, $ap\_r\_8-ap\_z\_8$,  $ap\_g\_4-ap\_g\_5$,
$ap\_z\_3-ap\_z\_4$, $ap\_r\_2-ap\_r\_3$, $ap\_r\_3-ap\_r\_4$, $ap\_g\_5-ap\_g\_6$,
$ap\_g\_4-ap\_r\_4$, $ap\_g\_6-ap\_g\_7$, $ap\_g\_1-ap\_g\_2$, $ap\_g\_3-ap\_g\_4$,
$ap\_g\_7-ap\_r\_7$, $ap\_z\_5-ap\_z\_6$, $ap\_g\_8-ap\_r\_8$, $ap\_z\_7-ap\_z\_8$, $ap\_z\_6-ap\_z\_7$,  $ap\_g\_2-ap\_g\_3$, $ap\_r\_3-ap\_z\_3$, $ap\_g\_7-ap\_g\_8$, $ap\_r\_6-ap\_r\_7$, $ap\_g\_3-ap\_r\_3$, $ap\_r\_1-ap\_z\_1$,  $ap\_r\_7-ap\_r\_8$,  $ap\_g\_1-ap\_r\_1$, $ap\_r\_5-ap\_r\_6$, $ap\_g\_2-ap\_r\_2$, $ap\_r\_2-ap\_z\_2$,
$ap\_z\_5-ap\_W1\_5$, $ap\_z\_2-ap\_W1\_2$,
$ap\_W1\_5-ap\_W2\_5$, $ap\_W1\_4-ap\_W2\_4$,
$ap\_W1\_2-ap\_W2\_2$, $ap\_W1\_1-ap\_W2\_1$, $ap\_W1\_2-ap\_W1\_3$,  $ap\_z\_4-ap\_W1\_4$, $ap\_z\_1-ap\_W1\_1$, $ap\_W1\_3-ap\_W2\_3$, $W1-ap\_W1\_1$, $W2-ap\_W2\_1$, $ap\_W1\_1-ap\_W1\_2$, $ap\_W2\_4-ap\_W2\_5$, $ap\_W2\_3-ap\_W2\_4$,
$ap\_W2\_2-ap\_W2\_3$, $ap\_W2\_1-ap\_W2\_2$, $ap\_W1\_4-ap\_W1\_5$, $ap\_W1\_3-ap\_W1\_4$, $ap\_z\_3-ap\_W1\_3$. (60 features).

Pattern III is $r - W1$, $g - r$, $g$, $r$, $r - z$, $W1 - W2$, $ap\_r\_4-ap\_z\_4$, $ap\_g\_5-ap\_r\_5$, $z-ap\_z\_1$, $W1$,  $r - W2$, $ap\_r\_6-ap\_z\_6$, $z$, $g-ap\_g\_1$, $ap\_z\_4-ap\_z\_5$, $r-ap\_r\_1$, $ap\_z\_1-ap\_z\_2$, $g-W1$, $ap\_r\_1-ap\_r\_2$, $g-z$,  $ap\_r\_4-ap\_r\_5$, $ap\_g\_6-ap\_r\_6$, $ap\_r\_7-ap\_z\_7$, $ap\_z\_2-ap\_z\_3$, $ap\_r\_5-ap\_z\_5$, $ap\_r\_8-ap\_z\_8$,  $ap\_g\_4-ap\_g\_5$, $W2$, $g-W2$,
$ap\_z\_3-ap\_z\_4$, $ap\_r\_2-ap\_r\_3$, $ap\_r\_3-ap\_r\_4$, $ap\_g\_5-ap\_g\_6$,
$ap\_g\_4-ap\_r\_4$, $ap\_g\_6-ap\_g\_7$, $ap\_g\_1-ap\_g\_2$, $ap\_g\_3-ap\_g\_4$,
$ap\_g\_7-ap\_r\_7$, $ap\_z\_5-ap\_z\_6$, $ap\_g\_8-ap\_r\_8$, $ap\_z\_7-ap\_z\_8$, $ap\_z\_6-ap\_z\_7$,  $ap\_g\_2-ap\_g\_3$, $ap\_r\_3-ap\_z\_3$, $ap\_g\_7-ap\_g\_8$, $ap\_r\_6-ap\_r\_7$, $ap\_g\_3-ap\_r\_3$, $ap\_r\_1-ap\_z\_1$,  $ap\_r\_7-ap\_r\_8$,  $ap\_g\_1-ap\_r\_1$, $ap\_r\_5-ap\_r\_6$, $ap\_g\_2-ap\_r\_2$, $ap\_r\_2-ap\_z\_2$,
(53 features).

\begin{thebibliography}{}
	\bibitem[Almosallam et~al. (2016)]{almosallam2016}Almosallam I. A., Lindsay S. N., Jarvis M. J., Roberts, S. J., 2016, MNRAS, 455, 2387
	\bibitem[Amaro et~al. (2019)]{amaro2019}Amaro V. et~al., 2019, MNRAS, 482, 3116
	\bibitem[Arnouts et~al. (1999)]{arnouts1999}Arnouts S., Cristiani S., Moscardini L., Matarrese S. Lucchin F. Fontana A. Giallongo E., 1999, MNRAS, 310, 540
	%\bibitem[Baum et~al., (1957)]{Baum57}Baum W. A., 1957, AJ. 62, 6
	\bibitem[Babbedge et~al. (2004)]{babbedge2004}Babbedge T. S. R. et~al., 2004, MNRAS, 353, 654
	\bibitem[Baldry et~al. (2018)]{gama2018}Baldry I. K. et~al., 2018, MNRAS, 474, 3875
%	\bibitem[Ball et~al. (2007)]{ball07}Ball N. M., Brunner R. J., Myers A. D., Strand N. E., Alberts S. L., Tcheng D., Llor$\grave{a}$ X., 2007, ApJ, 663(2), 774	
	\bibitem[Beck et~al. (2017)]{beck17}Beck R. et~al., 2017, MNRAS, 468, 4323
		\bibitem[Bender et~al. (2001)]{bender2001}Bender R. et~al., 2001, Deep Fields: Proceedings of the ESO Astrophysics symposia held at Garching, Germany, 9-12 Oct. 2000, Edited by S. Cristiani, A. Renzini, and R. E. Williams. Springer-Verlag, 2001, p.96
	\bibitem[Benitez et~al. (2000)]{benitez2000}Benitez N., 2000, ApJ, 536(2), 571
	%\bibitem[Blanton et~al., (2017)]{Blan17}Blanton M.~R. et~al., 2017, AJ, 154, 28
	\bibitem[Bonfield et~al. (2010)]{bon10}Bonfield D. G., Sun Y., Davey N., Jarvis M. J., Abdalla F. B., Banerji M., Adams R. G., 2010, MNRAS, 405(2), 987
	\bibitem[Bolzonella et~al. (2000)]{bolzonella2000}Bolzonelloa M., et~al., 2000, A\&A, 363, 476
	\bibitem[Brammer et~al. (2008)]{eazy2008}Brammer G. B., Van Dokkum P. G., Coppi P., 2008, ApJ, 686, 1503
	\bibitem[Breiman, (2001)]{Bre01}Breiman L., 2001, Machine Learning, 45, 5
	\bibitem[Brescia et~al. (2013)]{bre13}Brescia M., Cavuoti S., D'Abrusco R., Longo G., Mercurio A., 2013, ApJ, 772(2), article id: 140, 12 pp.
	\bibitem[Carliles et~al. (2010)]{car10}Carliles S., Budav$\acute{a}$ri T., Heinis S., Priebe C., Szalay A. S., 2010, ApJ, 712(1), 511
	%\bibitem[Carrasco \& Brunner (2014)]{car14}Carrasco K. M., Brunner, R. J., 2014, MNRAS, 438(4), 3409
	%\bibitem[Cavuoti et~al., (2012)]{cav12}Cavuoti S., Brescia M., Longo G., Mercurio A., 2012, A\&A, 546, id.A13, 8 pp
	%\bibitem[Cavuoti et~al., (2017)]{cav17}Cavuoti S., Amaro V., Brescia M., Vellucci C., Tortora C., Longo G., 2017, MNRAS, 465(2), 1959
	\bibitem[Chambers et~al. (2016)]{panstar2016} Chambers K. et~al., 2016, e-prints arXiv: 1612.05560
	\bibitem[Chen \& Guestrin, (2016)]{Chen2016}Chen T., Guestrin C., 2016, Proceedings of the 22nd ACM SIGKDD International Conference on Knowledge Discovery and Data Mining. ACM
	\bibitem[Connolly et~al. (1995)]{connolly1995}Connolly A. J., Csabai I., Szalay A. S., Koo D. C., Kron R. G., Munn J. A., 1995, AJ, 110, 2655
	\bibitem[Collister et~al. (2007)]{collister2007}Collister A. et~al., 2007, MNRAS, 375, 68
	\bibitem[Csabai et~al. (2003)]{csabai2003}Csabai I. et~al. 2003, AJ, 119, 69
	\bibitem[Chen et~al. (2018)]{chen2018}Chen C.-T. J. et~al., 2018, MNRAS, 478, 2132
	\bibitem[Cui et~al. (2012)]{Cui12}Cui X.-Q. et~al., 2012, RAA, 12, 1197
	\bibitem[Curran et~al. (2021)]{Curran2021}Curran S. J., Moss J. P., Perrott Y. C., 2021, MNRAS, 503, 2639
	\bibitem[de Jong et~al. (2013)]{kilo2013}de Jong J. T. A. et~al. 2013, The Messenger, 154, 44
	\bibitem[The DES Collaboration, (2005)]{DES2005}The Dark Energy Survey Collaboration, 2005, e-prints arXiv: astro-ph/0510346
	\bibitem[DESI Collaboration, (2016a)]{DESI2016a}DESI Collaboration, 2016a, e-prints arXiv: 1611.00037
	\bibitem[DESI Collaboration, (2016b)]{DESI2016b}DESI Collaboration, 2016b, e-prints arXiv: 1611.00036
	\bibitem[Desprez et~al. (2020)]{desprez20}Desprez G. et~al., 2020, A\&A, 644, A31
	\bibitem[Dey et~al. (2019)]{Dey2019}Dey A. et~al., 2019, AJ, 157(5), article id: 168
    \bibitem[Dorogush et~al. (2018)]{catboost2018}Dorogush A. V., Ershov V., Yandex A. G., 2018, e-prints arXiv:1810.11363
	\bibitem[Feldmann, et~al. (2006)]{feldmann2006}Feldmann R. et~al., 2006, MNRAS, 372, 565
	%\bibitem[Firth, Lahav \& Somerville, (2003)]{fir03}Firth A. E., Lahav O., Somerville R. S., 2003, MNRAS, 339(4), 1195
	\bibitem[Friedman et~al. (2001)]{Friedman2001}Friedman J. H., 2001, Annals of statistics, 29, 1189
	\bibitem[Gerdes et~al. (2010)]{Gerdes2010}Gerdes D. W., Sypniewski, A. J., Mckay, T. A., Hao, Jiangang, Weis, M. R., Wechsler, R. H., Busha, M. T., 2010, AJ, 715(2), 823
	\bibitem[Henghes et~al. (2021)]{Henghes2021}Henghes B., Pettitt C., Thiyagalingam J., Hey T., Lahav O., 2021, MNRAS, 505, 4847
	%\bibitem[Han et~al., (2021)]{Han2021}Han B., Qiao L.-N., Chen J.-L., Zhang X.-D., Zhang Y.-X., Zhao Y.-H., 2021, RAA, 219(1), 17
	%\bibitem[Hoyle (2016)]{hoy16}Hoyle B., 2016, Astronomy and Computing, 16, 34
	\bibitem[Jin et~al. (2019)]{Jinx2019}Jin X., Zhang Y., Zhang J., Zhao Y., Wu X., Fan, D., 2019, MNRAS, 485, 4539
	\bibitem[Jones \& Singal, 2017]{jon17}Jones E., Singal J., 2017, A\&A, 600, id.A113, 11 pp	
	%\bibitem[Koo et~al., (1985)]{Koo1985}Koo D. C., 1985, AJ, 90, 418
	\bibitem[Lang, Hogg \& Mykytyn, (2016)]{lang2016}Lang D., Hogg D. W., Mykytyn D., 2016, Astrophysics Source Code Library, record ascl:1604.008
	%\bibitem[Leistedt \& Hogg, (2017)]{lei17}Leistedt B., Hogg D. W., 2017, ApJ, 838(1), article id. 5, 14 pp
	\bibitem[Le Borgne \& Rocca-Volmerange, (2002)]{le2002}Le Borgne D., Rocca-Volmerange B., 2002, A\&A, 386, 446
	\bibitem[Li et~al. (2017)]{Li2017}Li C. et~al., 2017, Astroinformatics, Proceedings of the International Astronomical Union, IAU Symposium, 325, 353
	\bibitem[Li et~al. (2022)]{Li2022}Li C. et~al., 2022, MNRAS, 509, 2289
	\bibitem[Luo et~al. (2015)]{Luo15}Luo A. L. et~al., 2015, RAA, 15, 1095
	\bibitem[Lyke et~al. (2020)]{lyke20}Lyke B. W. et~al. 2020, ApJS, 250, 8
	\bibitem[Nishizawa et~al. (2020)]{nishizawa2020}Nishizawa, A. J., Hsieh, B. C., Tanaka, M., Takata, T., 2020, e-prints arXiv: 2003.01511
	\bibitem[Parkinson et~al. (2012)]{Parkinson2012}Parkinson D. et~al., 2012, Phys. Rev. D, 86
	\bibitem[Puschell et~al. (1982)]{Puschell1982}Puschell J., Owen F., Laing R. 1982, Proceedings of the International Astronomical Union, IAU Symposium, 97, 423
	\bibitem[Pedregosa et~al. (2011)]{scikit-learn}Pedregosa F. et~al., 2011, Journal of Machine Learning Research, 12, 2825
	\bibitem[Reis et~al. (2012)]{reis2012}Reis R. R. R. et~al., 2012, ApJ, 747, 59
	\bibitem[Ruknick et~al. (2001)]{ruknick2001}Rudnick G., et~al., 2001, AJ, 122, 2205
	\bibitem[Ruknick et~al. (2003)]{ruknick2003}Rudnick G., et~al., 2003, ApJS, 172, 117
	\bibitem[Sanchez et~al. (2014)]{sanchez14}S$\acute{a}$nchez C., et~al, 2014, MNRAS, 445(2), 1482
	\bibitem[Schmidt et~al. (2020)]{schmidt20}Schmidt S. J., et~al, 2020, MNRAS, 499, 1587
	\bibitem[Schindler et~al. (2017)]{sch17}Schindler J., Fan X., McGreer I. D., Yang Q., Wu J., Jiang L., Green R., 2017, ApJ, 851, 13
	\bibitem[Shuntov et~al. (2020)]{shuntov2020}Shuntov M., Pasquet J., Arnouts S., et~al., 2020, A\&A, 636, A90
	\bibitem[Wang et~al. (2007)]{wang07}Wang D., Zhang Y., Liu C., Zhao Y., 2007, MNRAS, 382(4), 1601
	\bibitem[Way \& Srivastava, (2006)]{way06} Way M. J., Srivastava A. N., 2006, ApJ, 647(1), 102	
	\bibitem[Way et~al. (2009)]{way09} Way M. J., Foster L. V., Gazis P. R., Srivastava A. N., 2009, ApJ,  706(1), 623
     \bibitem[York et~al. (2000)]{york00}York D. G. et~al., 2000, AJ, 120, 1579
    \bibitem[Yang et~al. (2014)]{yang2014}Yang G. et~al., 2014, ApJS, 215, 27
	\bibitem[Zhang et~al. (2013)]{zhang13}Zhang Y., Ma H., Peng N., Zhao Y., Wu X.-b., 2013, AJ, 146(2), article id. 22, 10pp
	\bibitem[Zhang et~al. (2019)]{zhang2019}Zhang Y.-X., Zhang J.-Y., Jin X., Zhao Y.-H., 2019, RAA, 19, 169
	\bibitem[Zhou et~al. (2021)]{zhou2021}Zhou R. et~al., 2021, MNRAS, 501, 3309
	\bibitem[Zou et~al. (2022)]{Huzou2022}Zou H. et~al., 2022, RAA, 22, 065001
\end{thebibliography}
\end{document}